\documentclass[
    twocolumn,
	prd,
	amssymb,
	preprintnumbers,
	secnumarabic,
	nofootinbib,
	superscriptaddress,
	]{revtex4-1}

\pdfoutput=1

\usepackage{feynmf}
\usepackage{graphicx}
\usepackage{enumitem}
\usepackage{latexsym}
\usepackage{amsfonts}
\usepackage{amssymb}
\usepackage{color}
\usepackage{amsmath}
\usepackage{slashed}
\usepackage{dcolumn}
\usepackage{verbatim}
\usepackage{comment}
\usepackage{float}
\usepackage{multirow}
\usepackage{xspace}
\usepackage[normalem]{ulem}
\usepackage{subcaption}
\usepackage[
pdfauthor={Nirmal Raj}]{hyperref}

\newcommand{\gsim}{\gtrsim}
\newcommand{\lsim}{\lesssim}

\def\lag{\mathcal{L}}
\def\Oc{\mathcal{O}}

\renewcommand{\tilde}{\widetilde} 

\newcommand{\beq}{\begin{eqnarray}}
\newcommand{\eeq}{\end{eqnarray}}
\newcommand{\bea}{\begin{eqnarray}}
\newcommand{\eea}{\end{eqnarray}}
\newcommand{\nn}{\nonumber}
\newcommand{\nnmb}{\nonumber}
\newcommand{\del}{\partial}
\newcommand{\kev}{\text{keV}}
\newcommand{\mev}{\text{MeV}}
\newcommand{\gev}{\text{GeV}}
\newcommand{\tev}{\text{TeV}}
\newcommand{\lrf}[2]{\left(\frac{#1}{#2}\right)}

\newcommand{\thdma}{2HDM+$a$ }

\def\RNS{R_{\rm NS}}
\def\MNS{M_{\rm NS}}
\def\mdm{m_{\chi}}

\def\mneff{\bar m_n}
\def\sigmageom{\sigma_0}
\def\Er{E_{\rm rec}}

\allowdisplaybreaks

\begin{document}

\title{Neutron star observations of pseudoscalar-mediated dark matter}

\author{John Coffey}
\email{jwcoffey@uvic.ca}
\affiliation{Department of Physics and Astronomy, University of Victoria, Victoria, BC V8P 5C2, Canada}

\author{David McKeen}
\email{mckeen@triumf.ca}
\affiliation{TRIUMF, 4004 Wesbrook Mall, Vancouver, BC V6T 2A3, Canada}

\author{David E. Morrissey}
\email{dmorri@triumf.ca}
\affiliation{TRIUMF, 4004 Wesbrook Mall, Vancouver, BC V6T 2A3, Canada}

\author{Nirmal Raj}
\email{nraj@triumf.ca}
\affiliation{TRIUMF, 4004 Wesbrook Mall, Vancouver, BC V6T 2A3, Canada}

\begin{abstract}
Scattering interactions between dark matter and Standard Model states mediated by pseudoscalars are generically challenging to uncover at direct detection experiments due to rates suppressed by powers of the local dark matter velocity $v_{\rm DM} \sim 10^{-3}\,c$.
However, they may be observed in the dark matter-induced heating of neutron stars, whose steep gravitational potentials prevent such suppression by accelerating infalling particles to semi-relativistic speeds.
We investigate this phenomenon in the context of two specific, self-consistent scenarios for pseudoscalars coupled to dark matter, and compare the sensitivity of neutron star heating to bounds from direct searches for the mediators and dark matter.
The first ``lighter" scenario consists of sub-10~GeV mass dark matter mediated by an axion-like particle~(ALP), while the second ``heavier" scenario has dark matter above $10\,\gev$ mediated by a dark pseudoscalar that mixes with a pseudoscalar from a two-Higgs doublet (the so-called \thdma model). 
In both frameworks, we show that imminent measurements of neutron stars will be able to test pseudoscalar-mediated
dark matter beyond the reach of direct dark matter searches as well as bounds on the mediators from flavor observables, beam dump experiments, and high-energy colliders.
\end{abstract}

\maketitle

\section{Introduction}
\label{sec:intro}

\begin{figure*}
    \centering
\makebox[\textwidth][c]{  \includegraphics[width=.49\textwidth]{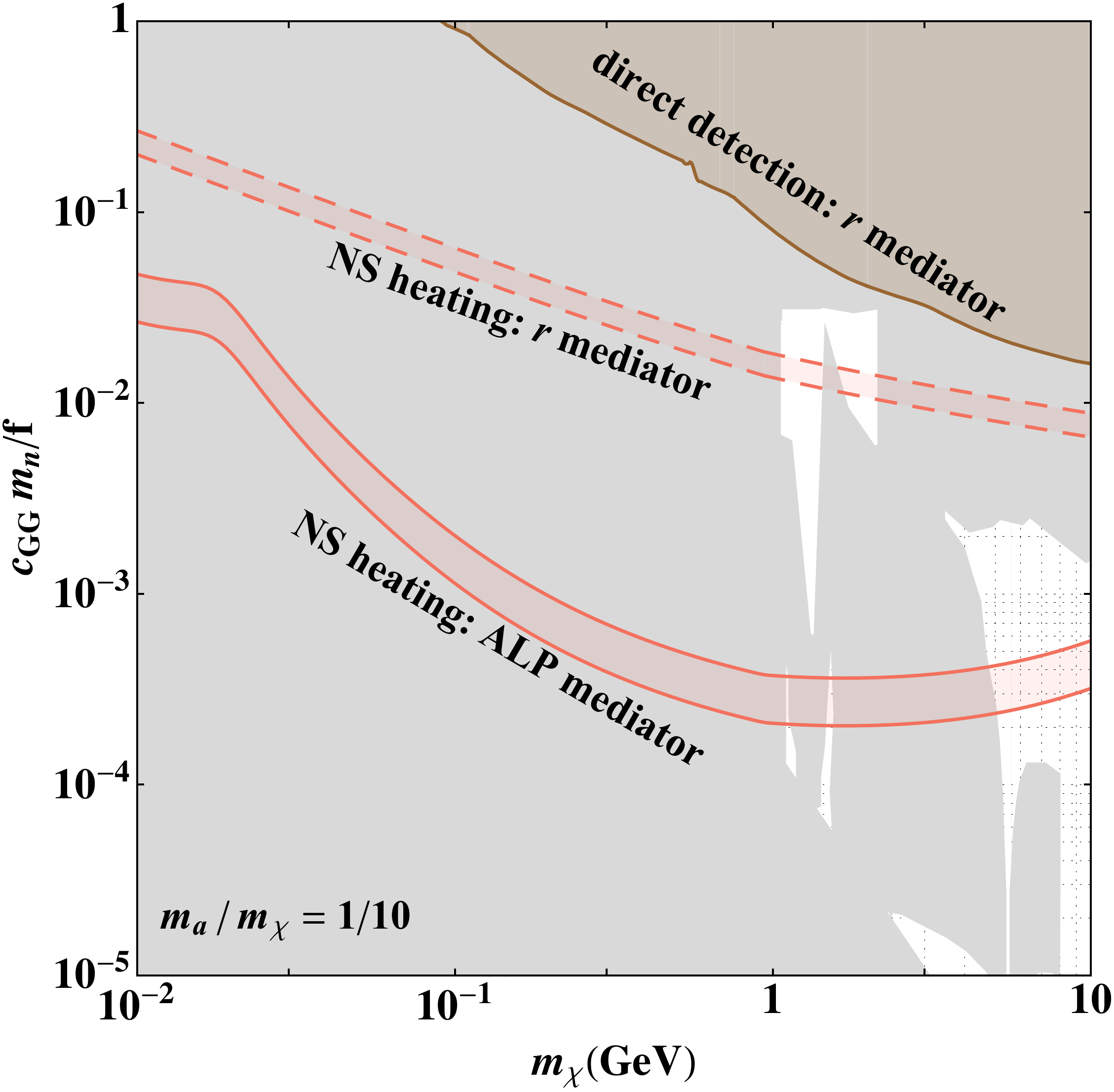} \
 \includegraphics[width=.49\textwidth]{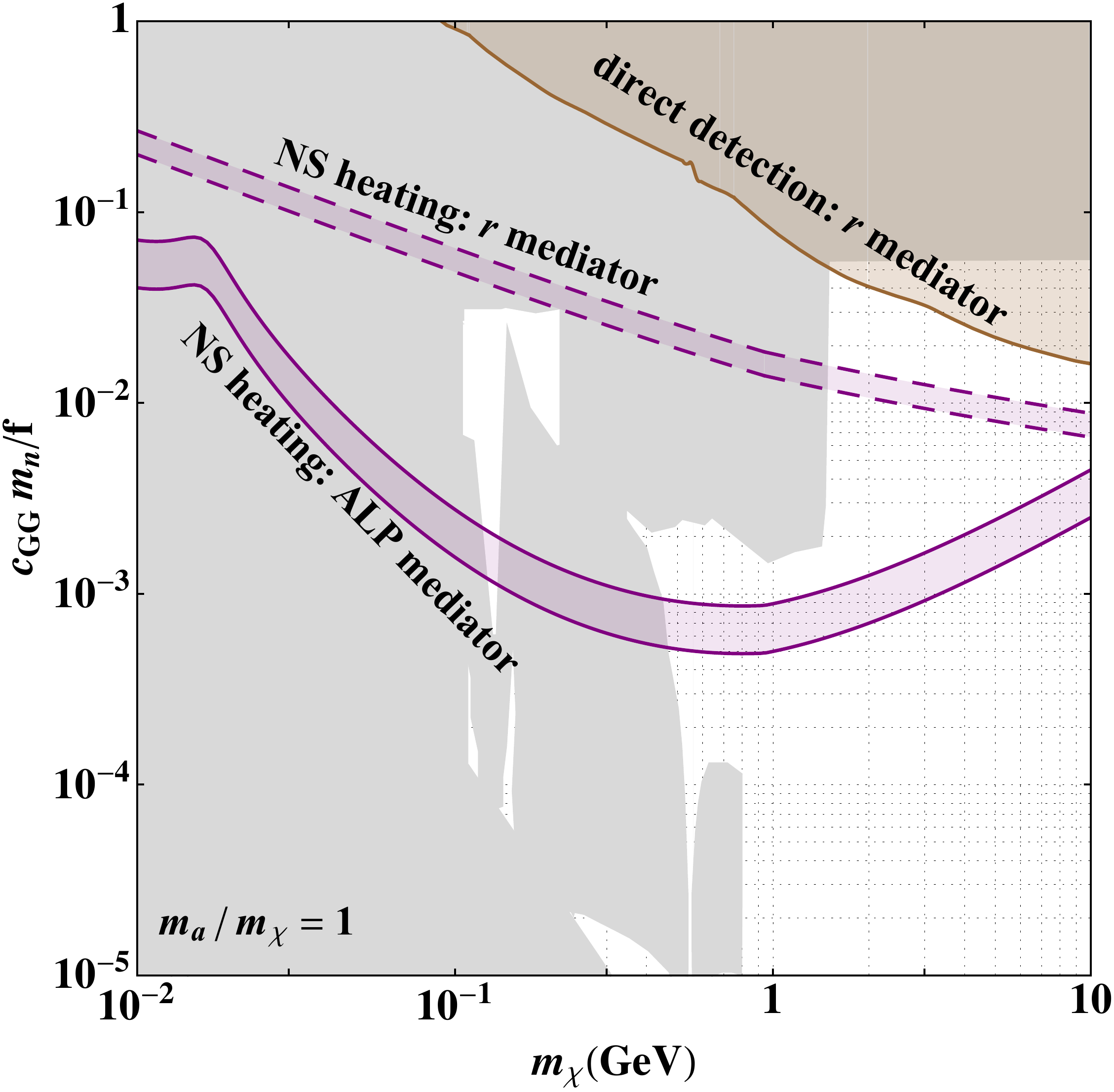} 
 }
\caption{
\raggedright{
\small Sensitivities to the ALP-gluon coupling versus dark matter mass, of neutron star heating from capture of DM mediated by the ALP ({\bf solid}) and the radial mode $r$ ({\bf dashed}) in our benchmark scenario, for ALP-to-DM mass ratios of 1/10 ({\bf left}) and 1 ({\bf right}). 
   The bands represent uncertainties in the mass-radius configurations of candidate neutron stars.
 Also shown are limits from dark matter direct searches ({\bf brown region}) and those on the mediator from beam dumps, rare meson decays, and astrophysics ({\bf gray region}).
   For further details, see Sec.~\ref{sec:results}. 
   }
      }
    \label{fig:lims-bauerBMcGG-masratios}
\end{figure*}


\begin{figure*}
    \centering
\makebox[\textwidth][c]{  \includegraphics[width=.49\textwidth]{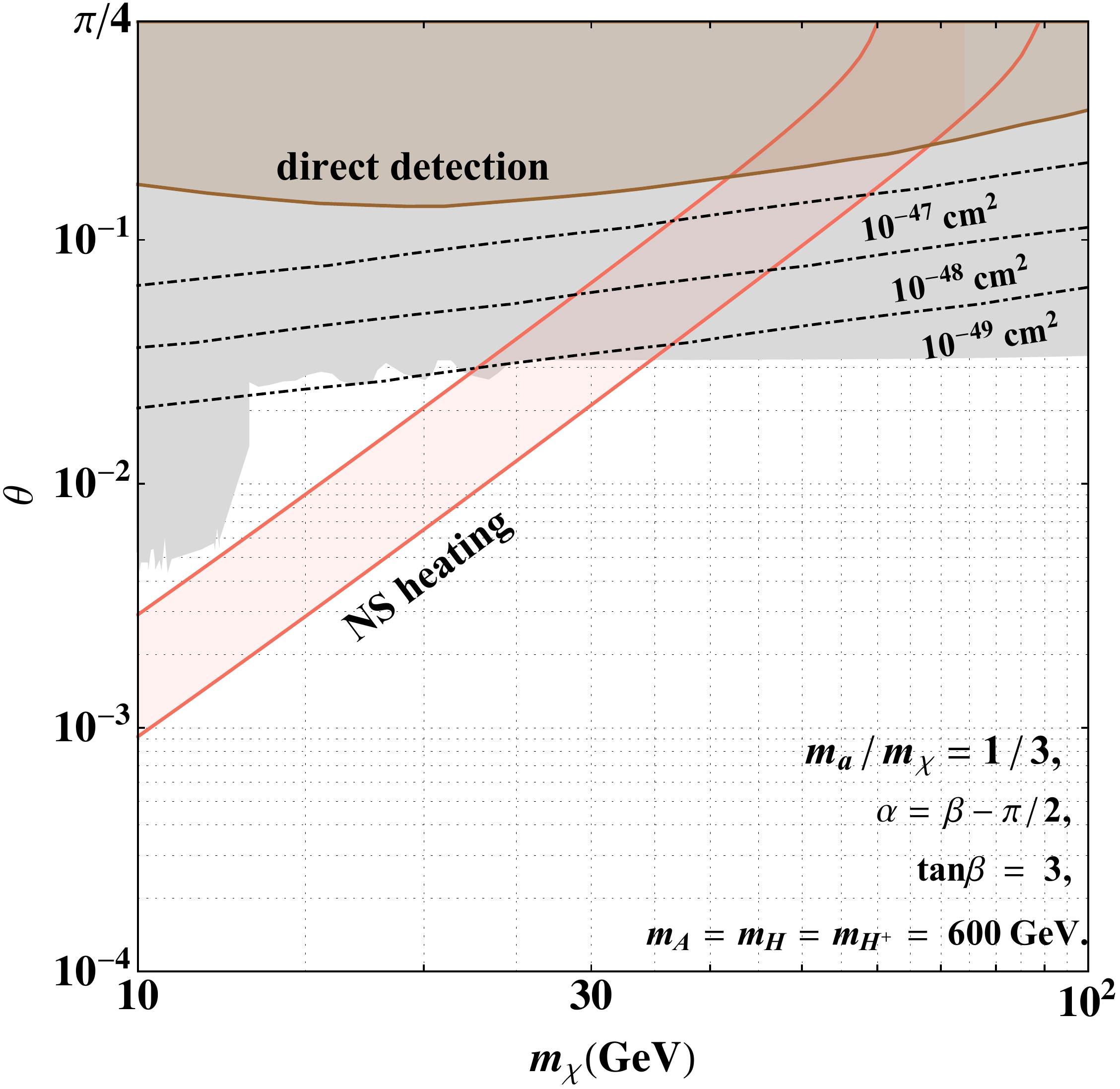} \   \includegraphics[width=.49\textwidth]{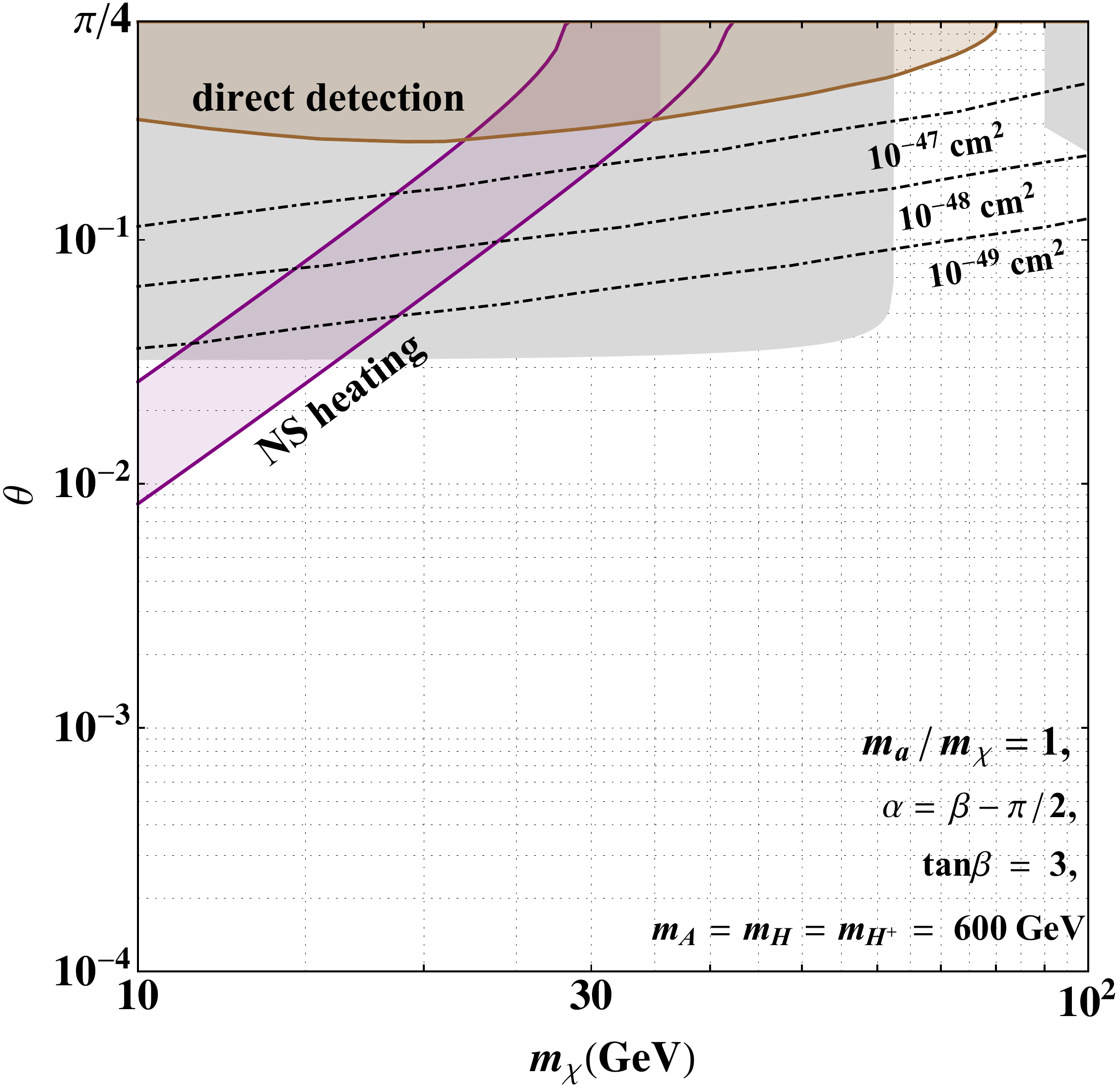} 
 }
 \caption{\raggedright{\small Sensitivities to the angle of mixing between the pseudoscalars in the 2HDM+$a$ scenario versus dark matter mass, of neutron star heating from capture of DM mediated by the pseudoscalar $a$ for our chosen benchmark point, for $a$-to-DM mass ratios of 1/3 ({\bf left}) and 1 ({\bf right}). 
 The bands represent uncertainties in the mass-radius configurations of candidate neutron stars.
 Also shown are limits from dark matter direct detection ({\bf brown region}), contours of cross sections within the reach of future detectors, and limits on the 2HDM+$a$ framework from colliders and flavor observables ({\bf gray region}).
 For further details, see Sec.~\ref{sec:results}.
 }
    }
    \label{fig:lims2HDMa}
\end{figure*}

Even as the hunt for dark matter (DM) is continuing on an unprecedented scale, its identity remains a closely guarded secret.
Direct searches for DM in the form of a weakly interacting massive particle~(WIMP)~\cite{Goodman:1984dc,Jungman:1995df} have placed remarkable limits on DM over the mass range of $m_\text{DM} \sim 1\,\gev$--$100\,\tev$~\cite{LUX:2016ggv,XENON:2018voc,PandaX-4T:2021bab}. The absence of WIMP signals in direct detection experiments has also motivated searches for other DM candidates over a broader range of masses~\cite{SnowmassLight:Mitridate:2022tnv,DEAPCollaboration:2021raj,SnowmassHeavy:Carney:2022gse}.
Notably, direct detection collaborations 
have made substantial progress in the ``light" DM regime of sub-GeV masses, where the lower target recoil energies make scattering more difficult to detect, with current limits down to $m_\text{DM} \sim 10\,\mev$ in nuclear recoils and $m_\text{DM} \sim 100\,\kev$ in electron recoils~\cite{reviewlightdm:KahnLin:2021ttr}.

Dark matter could also have avoided direct detection in other ways. 
For instance, should scattering of DM on targets proceed {\em in}elastically to an excited state (presumably present in the theory due to an approximate symmetry), the target recoil rates diminish rapidly with increasing mass splitting between the dark states~\cite{Tucker-Smith:2001myb,Tucker-Smith:2004mxa}.
Likewise, the scattering cross section of local DM in a target can be highly suppressed if the underlying dynamics make it proportional to positive powers of the DM velocity, $v_\text{DM} \sim 10^{-3}$~\cite{Fan:2010gt,Freytsis:2010ne,Fitzpatrick:2012ix,Dienes:2013xya}. 
This latter scenario is realized when DM connects to visible matter through a pseudoscalar mediator.
A comprehensive program to discover DM via scattering must necessarily include probes beyond direct detection.

In recent times, signals from DM capture in compact stars have emerged as a highly promising avenue; for reviews, see Refs.~\cite{review:Tinyakov:2021lnt,snowmassWP:Berti:2022rwn}.
In particular, capture in neutron stars~(NS), beginning with Ref.~\cite{Goldman:1989nd}, has attracted much attention by virtue of properties favorable to DM capture, namely their steep gravitational potentials and high densities.
One of us recently showed with other collaborators that old, isolated NSs would heat appreciably by the transfer of kinetic energy from infalling DM, and that the phenomenon could be detected with forthcoming infrared telescopes trained at nearby NSs~\cite{NSvIR:Baryakhtar:DKHNS}.
The observation of a population of anomalously warm NSs would therefore provide evidence for DM, while the discovery of even a single relatively cold NS could be used to exclude a broad range of DM candidates.
Much theoretical work has followed this proposal~\cite{NSMultiscat:Bramante:2017xlb,NSvIR:Raj:DKHNSOps,NSvIR:Pasta,NSvIR:SelfIntDM,NSvIR:Bell2018:Inelastic,NSvIR:GaraniGenoliniHambye,NSvIR:Queiroz:Spectroscopy,NSvIR:Hamaguchi:Rotochemical,NSvIR:Marfatia:DarkBaryon,NSvIR:Bell:Improved,NSvIR:DasguptaGuptaRay:LightMed,NSvIR:GaraniGuptaRaj:Thermalizn,NSvIR:Queiroz:BosonDM,NSvIR:Bell2020improved,NSvIR:zeng2021PNGBDM,NSvIR:anzuiniBell2021improved,NSvIR:Bell2019:Leptophilic,NSvIR:GaraniHeeck:Muophilic,NSvIR:Riverside:LeptophilicShort,NSvIR:Riverside:LeptophilicLong,NSvIR:Bell:ImprovedLepton,NSvIR:HamaguchiEWmultiplet:2022uiq,NsvIR:HamaguchiMug-2:2022wpz}, exploring several important particle and nuclear astrophysics aspects. 

NS heating by dark matter is also strongly complementary to other search methods.
In particular, this indirect probe is sensitive to many types of particle-like DM that are challenging to detect through direct scattering in the laboratory.
This includes lighter sub-GeV DM~\cite{NSvIR:Bell2019:Leptophilic,NSvIR:Riverside:LeptophilicShort,NSvIR:Riverside:LeptophilicLong,NSvIR:Pasta,NSvIR:Marfatia:DarkBaryon},
DM that scatters primarily inelastically~\cite{NSvIR:Baryakhtar:DKHNS,NSvIR:Bell2018:Inelastic,NSvIR:Pasta,NSvIR:HamaguchiEWmultiplet:2022uiq}, and scenarios where the scattering rate of DM with nuclei is suppressed by positive powers of the DM velocity~\cite{NSvIR:Raj:DKHNSOps,NSvIR:Bell2019:Leptophilic,NSvIR:Riverside:LeptophilicLong}. 

In this study, we examine the sensitivity of the NS kinetic heating mechanism to DM that connects to the SM through a pseudoscalar mediator. As mentioned above, pseudoscalar mediation suppresses scattering cross sections in direct detection by multiple powers of DM velocity.
Neutron star heating overcomes this suppression by virtue of the semi-relativistic speeds with which DM falls into NSs.
This fact has been noted before~(in, e.g. Refs.~\cite{NSvIR:Raj:DKHNSOps,NSvIR:Bell2019:Leptophilic,NSvIR:Riverside:LeptophilicLong}), but only in the context of a generic pseudoscalar in the ``heavy mediator" limit, {\em i.e.} in an effective field theory (EFT) regime where the mediator mass exceeds the momentum transfer. 
We expand on this previous work by investigating the sensitivity of NSs to pseudoscalar-mediated DM within self-consistent, ultraviolet~(UV) complete scenarios. 
This allows us to extend beyond the range of EFT validity to cases where the mediator is light relative to the relevant 4-momentum exchange, and to compare the reach of NS heating observations to other tests of the DM particle as well as the mediator itself.

To be concrete, in this work we study fermionic dark matter $\chi$ and two realizations of lighter pseudoscalar mediators. 
The first realization is in the form of an axion-like particle~(ALP) that couples primarily to gluons, for which a UV completion may be specified, and on which constraints have been placed by a multitude of experiments. 
Since the ALP can be naturally light relative to the weak scale, we focus on ALP mediators in the context of lighter DM with $m_\chi \lesssim 10\,\gev$. 
The second mediator realization we consider is the well-studied Two Higgs Doublet Model plus pseudoscalar, the 2HDM+$a$. 
Here, the lighter pseudoscalar relevant to DM detection couples primarily to quarks and has a mass that is typically near the weak scale. 
We focus on this scenario to study DM in the traditional WIMP mass range, $\mdm \gsim 10$~GeV.  

In both these scenarios we find regions where the discovery potential of NS heating is a considerable improvement over existing constraints.
Our primary results are summarized in Fig.~\ref{fig:lims-bauerBMcGG-masratios} for the ALP mediator and Fig.~\ref{fig:lims2HDMa} for the \thdma mediator.
ALPs are generically excluded by beam dump searches for masses below a few 100 MeV, but NS heating would probe surviving islands of parameter space in this region as well as higher masses.
The 2HDM+$a$ scenario for certain parametric ranges remarkably survives a host of collider and flavor constraints in the so-called alignment limit, and we identify the regions that NS heating can exclusively probe.

This paper is organized as follows. 
Following this introduction, we describe our benchmark ALP and \thdma scenarios in Section~\ref{sec:models}.
Next, in Section~\ref{sec:direct} we provide expressions for nucleon scattering cross sections in both scenarios.
In Section~\ref{sec:NScapture} we review neutron star capture of dark matter with an emphasis on kinetic heating, and discuss how these apply specifically to pseudoscalar mediation.
In Section~\ref{sec:results} we present our results.
Section~\ref{sec:concs} is reserved for conclusions and outlook.
Various technical details underlying our main results are presented in the Appendices~\ref{sec:appa}, \ref{sec:appb}, \ref{sec:appc}.

\section{Benchmark Scenarios}
\label{sec:models}

Our investigation focuses on Dirac fermion dark matter $\chi$ that connects to the SM primarily through a lighter pseudoscalar mediator.  We do not attempt to specify the cosmological origin of the $\chi$ abundance, but we note that it can be consistent with bounds from indirect detection if its abundance is 
asymmetric~\cite{Nussinov:1985xr,Barr:1990ca,Kaplan:2009ag}.
Furthermore, our assumption of a pseudoscalar mediator also prevents large non-perturbative enhancements of the annihilation and self-scattering cross sections, even when the mediator is parametrically lighter than the DM state~\cite{Agrawal:2020lea}.

In our study we consider two benchmark scenarios for the pseudoscalar mediator. In the first,  we investigate light ({\em i.e.} mass well below the weak scale) mediators that can arise as axion-like particles~(ALPs) from the spontaneous breaking of an approximate global symmetry. In the second benchmark, we use the \thdma model in which a singlet pseudoscalar couples to dark matter directly and connects to the SM by mixing with a heavier pseudoscalar from a two-Higgs doublet~(2HDM) extension of the SM. This second scenario motivates weak-scale mediator masses. Thus, taking $m_a \leq m_\chi$, we focus on the ALP-mediator benchmark for lighter DM with $m_\chi \lesssim 10\,\gev$ and the \thdma-mediator benchmark for heavier DM with $m_\chi \gtrsim 10\,\gev$.

\subsection{Lighter Mediator: Axion-Like Particles}

A natural way to realize a light pseudoscalar below the weak scale is in the form of an axion-like particle~(ALP) that emerges as the pseudo-Nambu-Goldstone boson from the spontaneous breaking of an approximate but anomalous global $U(1)$ symmetry~\cite{Peccei:1977hh,Peccei:1977ur,Weinberg:1977ma,Wilczek:1977pj}. 
Its mass is set by either the soft breaking scale of the $U(1)$ or the mass scale of the gauge sector containing the anomaly, and can be parametrically small. Our first benchmark for a pseudoscalar mediator for DM consists of an ALP that couples primarily to gluons.

\subsubsection{Model and Elementary Couplings}

In this context we consider a light pseudoscalar arising as the phase of a complex scalar $\phi$ that couples to DM according to
\beq
\lag \ \supset \ - g_\chi\,\phi\,\bar{\chi}_R\chi_L + {\rm h.c.} 
\label{eq:alpdm1}
\eeq
This is consistent with a global $U(1)_{PQ}$ symmetry under which $[\phi]_{PQ}=1$ and $[\chi_L]_{PQ}-[\chi_R]_{PQ}=-1$. If $\phi$ develops a vacuum expection value~(VEV), we can write 
\beq
\phi = \frac{f}{\sqrt{2}}\,(1 + r/f)\,e^{ia/f} \, 
\eeq
with the real scalar $r(x)$ and pseudoscalar ALP $a(x)$. This leads to a mass for $\chi$,
\beq
m_\chi = g_\chi\,f/\sqrt{2} \ .
\eeq
Changing variables such that $\chi_L \to e^{-i\beta_L\,a/f}\chi_L$ and $\chi_R\to e^{-i\beta_R\,a/f}\chi_R$ with $(\beta_L-\beta_R) = 1$ removes the pseudoscalar from Eq.~\eqref{eq:alpdm1} at the cost of introducing new interactions~\cite{DiLuzio:2020wdo}:
\beq
\lag  &\supset& 
\frac{\del_\mu a}{f}\,
\big[ 
\beta_L\,\bar{\chi}_L\gamma^{\mu}\chi_L + \beta_R\,\bar{\chi}_R\gamma^{\mu}\chi_R 
\big]
\label{eq:alpdm2}\\
&=& \frac{\del_\mu a}{2f}\,
\big[(\beta_L+\beta_R)\,
\bar{\chi}\gamma^{\mu}\chi - \bar{\chi}\gamma^{\mu}\gamma^5\chi
\big]\nnmb\\
&=&  (\text{E.O.M.})
+ {m_\chi}\,\frac{a}{f}\;\bar{\chi}i\gamma^5\chi
 \ ,
\nnmb
\eeq
where (E.O.M.) refers to terms that vanish on-shell when the equations of motion are satisfied. We also include an explicit mass term for the ALP $a(x)$ that softly breaks the global $U(1)$ and allows us to treat the ALP mass $m_a$ as a free parameter.

The pseudoscalar ALP can connect to the SM in the presence of fermions with SM gauge charges that also get their mass from the VEV of $\phi$. 
To be concrete, we assume a set of $N_Q$ fermions $Q$ with SM charges $({\bf 3,1,0})$ together with the coupling~\cite{Kim:1979if,Shifman:1979if}
\beq
\lag \supset - y_{Q}\,\phi\,\bar{Q}_{R}{Q}_L + {\rm h.c.}
\eeq
As before, the interaction generates fermion masses $m_Q = y_Q\,f/\sqrt{2}$ as well as couplings of the form of Eq.~\eqref{eq:alpdm2} after shifting field variables. The change of field variables also generates an interaction between the pseudoscalar and gluons~\cite{Bauer:2017ris,Bauer:2020jbp}:
\beq
\lag  &\supset&  \frac{\alpha_s}{4\pi}\,\frac{N_Q}{2}\,\frac{a}{f}\,G_{\mu\nu}^a\widetilde{G}^{a\,\mu\nu} 
 \label{eq:alpglu}\\
 &\equiv&
 c_{GG}\,\frac{\alpha_s}{4\pi}\,\frac{a}{f}\,G_{\mu\nu}^a\widetilde{G}^{a\,\mu\nu} \ ,
 \nnmb
\eeq
with $c_{GG} = N_Q/2$.
We note that the ALP literature often expresses this coupling in terms of $f_a = f/2c_{GG}$~\cite{DiLuzio:2020wdo}.
The interaction of Eq.~\eqref{eq:alpglu} is the primary connection between the ALP and the SM.

Additional couplings to the SM will be generated by running from the matching scale $f$ down to the energies relevant for DM scattering. Below the electroweak scale, these take the form~\cite{Bauer:2017ris,Bauer:2020jbp,Bauer:2021wjo,Bauer:2021mvw}
\beq
\lag \supset \frac{\del_\mu a}{f}\big(
\bar{f}_L\gamma^{\mu}\mathbf{k}_Ff_L + \bar{f}_R\gamma^{\mu}\mathbf{k_f}f_R
\big)
\label{eq:alpferm}
\eeq
where $f_{L,R}$ refer to SM fermions and $\mathbf{k_F}$ and $\mathbf{k}_f$ are matrices in flavor space. For the dark matter analysis to follow, we will be interested mainly in the flavor-diagonal coupling combinations $c_{ii} = (k_{f_{ii}}-k_{F_{ii}})$. Relative to $c_{GG}$ defined in Eq.~\eqref{eq:alpglu}, we find $c_{ii} \sim 10^{-2}c_{GG}$.

Together with the ALP, this model also contains the radial mode $r(x)$ of the complex scalar, $\phi(x) = (f/\sqrt{2})(1+r/f)e^{ia/f}$. In contrast to the ALP, however, this state obtains a mass directly from spontaneous symmetry breaking with $m_r \propto f$. The most important couplings of $r$ for our analysis are
\beq
\lag \ &\supset& \
\frac{r}{f}\,(\del a)^2
-m_\chi\,\frac{r}{f}\,\bar{\chi}\,\chi
\label{eq:radialglu}\\
&&~~~
+\frac{\alpha_s}{6\pi}\,c_{GG}\,\frac{r}{f}\,G_{\mu\nu}^aG^{a\,\mu\nu} \ ,
\nnmb
\eeq
where the last term is generated by integrating out the heavy color-charged fermions for $m_Q\gg m_r$.

\subsubsection{Limits on the ALP Model}

The most stringent bounds on the theory typically come from the ALP mediator itself. Experimental limits on ALP-type pseudoscalars coupling to the SM primarily through the operator of Eq.~\eqref{eq:alpglu} have been studied recently in Refs.~\cite{Bauer:2017ris,Bauer:2020jbp,Bauer:2021wjo,Bauer:2021mvw}, while astrophysical limits from supernova physics are calculated in Refs.~\cite{ErtasKahlhoefer:2020xcc,Caputo:2022mah,Caputo:2021rux}. We review and summarize these limits in App.~\ref{sec:appa} and show the exclusions as shaded regions in our sensitivity plots.

In addition to the ALP, the theory also contains the heavier radial mode $r$ and mediator quarks $Q$ with masses near $m \sim f$. These are discussed in App.~\ref{sec:appa} and can lead to further bounds on the theory, particularly for ALP masses above $m_a \gtrsim 2\,\gev$. While the detailed bounds depend on the specific UV details of the model, we find that all current limits can be satisfied over the entire ALP mass region of interest for $c_{GG}/f \gtrsim 10^{-2}\,\gev^{-1}$.

\subsection{Heavier Mediator: \thdma}

As a benchmark model of a heavier pseudoscalar mediator ($m_a \gtrsim 10\,\gev$), we consider the well-studied \thdma theory consisting of a two-Higgs doublet model~(2HDM) together with an additional pseudoscalar $a_0$ that couples to fermionic dark matter $\chi$~\cite{Ipek:2014gua}.

\subsubsection{Model and Elementary Couplings}

In this model a gauge-singlet  pseudoscalar $a_0$ interacts with the Dirac fermion DM field $\chi$ via
\beq
\lag_\chi = y_\chi\,a_0\,\bar{\chi}i\gamma^5\chi \ .
\label{eq:axx}
\eeq
The $a_0$ also connects to the SM by coupling to the $H_1$ and $H_2$ doublets of the 2HDM sector according to
\beq
\lag_{a12} =  - i\,\mu_{a12}\, a_0\,H_1^{\dagger}H_2 + {\rm h.c.}
\label{eq:ahh}
\eeq
Together, the interactions of Eqs.~(\ref{eq:axx},\ref{eq:ahh}) connect the DM particle to the SM via the pseudoscalar $a_0$.

The scalar sector of the 2HDM+$a$ model is described by the potential 
\beq
V = V_\text{2HDM} + \frac{1}{2}m_{a_0}^2a_0^2 - \lag_{a12} \ ,
\eeq
where $V_\text{2HDM}$ is a standard two-Higgs doublet potential~\cite{Gunion:1989we,Djouadi:2005gj}. On its own, the 2HDM gives rise to physical scalars $h$ and $H$, a pseudoscalar $A_0$, and a charged complex scalar $H^{\pm}$. The 2HDM parameters in this potential can be defined in terms of the expectation values $v_1$ and $v_2$ of the doublets $H_1$ and $H_2$ with $v = \sqrt{v_1^2+v_2^2}=246\,\gev$, $\tan\beta = v_2/v_1$, the CP-even scalar mixing angle $\alpha$, the 2HDM masses $m_h$, $m_H$, $m_{A_0}$, $m_{H^\pm}$, and the quartic coupling $\lambda_3$ connecting $|H_1|^2$ and $|H_2|^2$. We assume further that the $H_1$ and $H_2$ doublets couple to the SM fermions in the Type-II format where $H_1$ connects only to $d$- and $e$-type fermions and $H_2$ couples exclusively to $u$-type fermions.
Including the term of Eq.~\eqref{eq:ahh} leads to mixing between the $a_0$ and $A_0$ states described by the angle
\beq
\tan 2\theta = - \frac{2 \mu_{a12} v}{m_A^2-m_a^2} \ ,
\eeq
where $m_a$ and $m_A$ are the physical masses of the lighter and heavier pseudoscalars. Together with the angle $\theta$, these three parameters can be used to specify the theory in place of $m_{a_0}$, $m_{A_0}$, and $\mu_{a12}$.

After electroweak symmetry breaking, 
the physical pseudoscalar mass eigenstates $a$ and $A$ couple to DM and SM fermions according to
\bea
\nn -\lag_{\rm Yuk} &\supset& \sum_f \frac{m_f}{v} (\xi_a^f a + \xi_A^f A) \bar f i \gamma^5 f  \\
    &&~~ + (\xi_a^\chi a + \xi_A^\chi A ) \bar \chi i \gamma^5 \chi
\eea 
where for a Type-II 2HDM,
\beq
\begin{array}{rclcrcl}
\xi_a^u &=& \sin \theta\,\cot \beta \ ,&~& \xi_a^d &=& \sin \theta\,\tan \beta \ ,\\
\xi_A^u &=& -\cos \theta\,\cot \beta \ ,&~&\xi_A^d &=& -\cos \theta\,\tan \beta \ , \\
\xi_a^\chi &=& y_\chi \cos \theta\ ,&~&\xi_A^\chi &=& y_\chi \sin \theta \ .
\end{array}
\label{eq:2HDMaYuk}
\eeq

\subsubsection{Limits on the \thdma Model}
\label{sec:2HDMa_limits}
For the dark matter studies to follow, we will focus on the model in the limit of $m_a \ll m_A$ and $|\theta|\ll 1$ in which the light pseudoscalar is mostly singlet. The bounds on the theory in this limit then mostly factorize into those on the 2HDM sector and those on the light pseudoscalar. 

As a fixed benchmark for the 2HDM part of the theory, we set $\tan\beta = 3$, $m_H = m_{H^\pm} = m_A = 600\,\gev$, $\lambda_3 \simeq m_h^2/v^2$, and work in the alignment/decoupling limit of $\cos(\beta-\alpha)=0$ such that the lighter $h$ scalar couples to the rest of the SM in the same way as the SM Higgs boson~\cite{Gunion:1989we,Djouadi:2005gj}. 
As discussed further in Appendix~\ref{sec:appb}, in the absence of pseudoscalar mixing these parameters are consistent with direct and indirect bounds on the Type-II 2HDM including flavour, precision electroweak and Higgs measurements, and collider searches. It also satisfies the requirements of vacuum stability and perturbative unitarity~\cite{Kling:2016opi}.

Within this 2HDM-sector benchmark we allow for variations in $m_a$ and $\theta$. 
We briefly summarize the bounds on this parameter subspace here, and discuss them in more detail in Appendix~\ref{sec:appb}. 

For $m_a \lesssim 10\,\gev$ flavor constraints arise.
For $1\,\gev\lesssim m_a \lesssim 10\,\gev$ the most important of these come from the loop-induced $b$-$s$-$a$ coupling~\cite{Arcadi:2017wqi,Arnan:2017lxi}. 
This effective coupling leads to rare decays of bottom mesons, such as $B_s\to\mu^+\mu^-$ and $B^\pm\to {K^\pm}^{(\ast)} \mu^+\mu^-$. 
Limits from the latter process~\cite{LHCb:2015nkv} constrain the mixing angle to be smaller than ${\cal O}({\rm few}\times10^{-3})$ for $m_a\lesssim 4\,\gev$.

When $m_a \leq m_h/2$ the decay channel $h\to aa$ can have a significant branching fraction for $|\theta| \gtrsim 0.05$ leading to bounds on $\theta$ of this order from searches for exotic Higgs boson decays as well as Higgs rate measurements. 
The strongest limits for larger $m_a\geq 90\,\gev$ come from LHC searches for $H/A\to \tau\tau$ and imply $|\theta| \lesssim 0.1$ up to $m_a \simeq 2\,m_t$ (for $\tan\beta = 3$). These bounds are shown in Fig.~\ref{fig:limsmed-2hdma} in App.~\ref{sec:appb} and are summarized by the shaded grey regions in Fig.~\ref{fig:lims2HDMa}.

\section{Dark Matter Direct Detection}
\label{sec:direct}

\begin{figure*}
    \centering
    \begin{subfigure}[t]{0.35\textwidth}
 \includegraphics[width=0.6\textwidth]{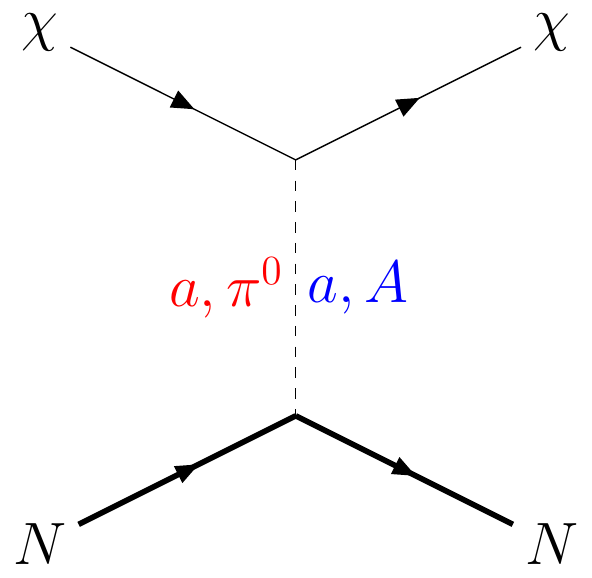} 
 \caption{{\bf spin- and velocity-dependent} via exchange of \textcolor{red}{ALP \& pion}, or \textcolor{blue}{2HDM+$a$ pseudoscalars}.}
 \end{subfigure} \ \ 
     \begin{subfigure}[t]{0.35\textwidth}
 \includegraphics[width=0.6\textwidth]{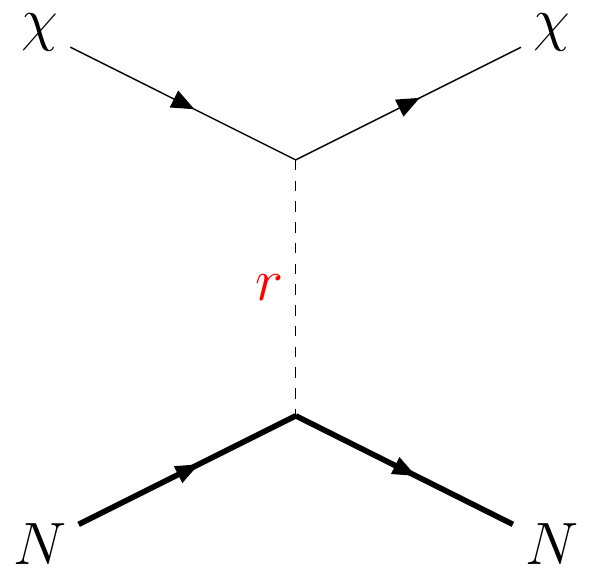} 
  \caption{{\bf spin- and velocity-independent} via exchange of \textcolor{red}{ALP radial mode}.}
 \end{subfigure} 
 \vskip \baselineskip
     \begin{subfigure}[t]{.7\textwidth}
     \centering
 \includegraphics[width=0.3\textwidth]{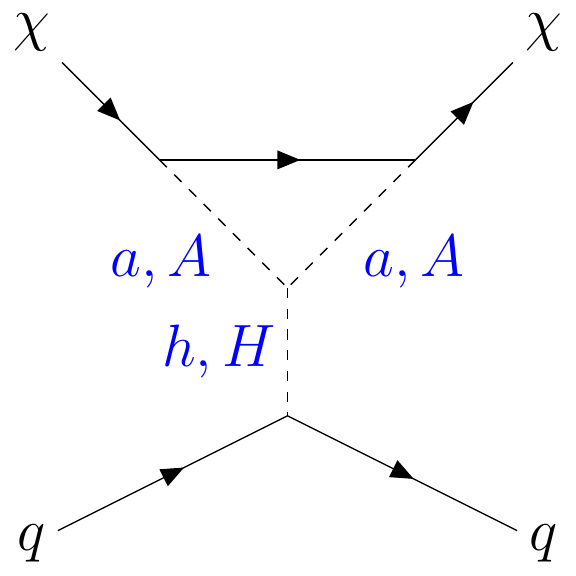} \ 
  \includegraphics[width=0.3\textwidth]{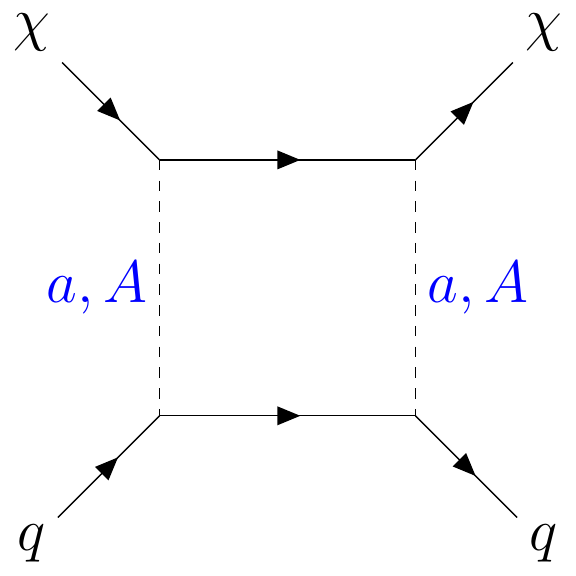}
  \caption{{\bf spin- and velocity-independent} loops involving \textcolor{blue}{2HDM+$a$ pseudoscalars}.}
 \end{subfigure}
 \caption{\raggedright{\small Feynman diagrams leading to dark matter scattering on nucleons in our setups. 
 The mediators involved in our 
 ALP and \thdma pseudoscalar scenarios
 are respectively labeled in \textcolor{red}{red} and \textcolor{blue}{blue}.}
 }
    \label{fig:feyndiag}
\end{figure*}

In this section we provide expressions for the non-relativistic nuclear scattering cross sections for DM direct detection in the two benchmark models presented above. Both models give rise to spin-dependent scattering on nuclei through direct pseudoscalar exchange as well as subdominant spin-independent scattering by other mechanisms. We present the leading contributions to each channel. 

\subsection{ALP Model Results}

The dominant contributions to spin-dependent and spin-independent scattering in the ALP scenario are shown in panels (a) and (b) of Fig.~\ref{fig:feyndiag}, respectively. 
The former arises from exchanging the pseudoscalar $a$ together with its mixing with the neutral pion, and the latter from the radial mode $r(x)$ in the $t$-channel. Both are dominated by the couplings of the scalars to gluons rather than to quarks.

\subsubsection{ALP Spin-Dependent}

The interactions of Eqs.~(\ref{eq:alpglu},\ref{eq:alpferm}) connect the axion to nucleons and lead to mixing with the neutral mesons. Thus, the axion serves as a mediator between DM and nucleons. To estimate these effects, we work in the limit of two quark flavors as in Refs.~\cite{Bauer:2021wjo,Bauer:2021mvw} (although see Ref.~\cite{Blinov:2021say} for a related analysis that includes three flavors). 
We estimate that an explicit three-flavor treatment leads to only small changes at low momentum exchange $Q \equiv \sqrt{|q^2|} \lesssim 500\,\mev$, while for higher momenta other uncertainties are expected to be more important. 

The mixing between the axion and the neutral pion in this context depends on the small parameter
\beq
\varepsilon \equiv \frac{f_\pi}{2f}\left[c_{uu}-c_{dd}+2c_{GG}\lrf{m_d-m_u}{m_u+m_d}\right] \ ,
\eeq
where $f_\pi \simeq 93\,\mev$ is the pion decay constant.
To leading non-trivial order in $\varepsilon$, axion-pion mixing is described by the mixing angle $\alpha$ given by
\beq
\tan(2\alpha) = -2\varepsilon\,\frac{\text{min}\{m_a^2,\,m_\pi^2\}}{m_\pi^2-m_a^2} \ ,
\eeq
with physical masses
\beq
m_{\bar{a},\bar{\pi}}^2 &=& \frac{1}{2}\bigg[(m_a^2+m_\pi^2) \mp\\ &&(m_\pi^2-m_a^2)\sqrt{1+\frac{4\varepsilon^2\,\text{min}\{m_\pi^4,m_a^4\}}{(m_\pi^2-m_a^2)^2}}\bigg]~,
\nnmb
\eeq
where
\beq
m_\pi^2 &=& B_0(m_u+m_d)~,\\
m_a^2 &=& m_{a_0}^2 + 4\,m_\pi^2\,\frac{m_um_d}{(m_u+m_d)^2}\lrf{f_\pi}{f}^2c_{GG}^2~.\nnmb
\eeq
In terms of these parameters, and working to linear order in $\varepsilon$ but to all orders in $\alpha$ (to accommodate $m_a^2\simeq m_\pi^2$), we find the physical axion and neutral pion couplings to nucleons $N=p,n$ and DM $\chi$ to be 
\beq
\lag &\supset&
\bar{N}\,\gamma^\mu\gamma^5\left(\frac{g_0}{2}u^{(0)}_\mu + \frac{g_A}{2}\sigma^3u^{(1)}_{\mu}\right)N
\nnmb\\
&&\ + \ \bar{\chi}\gamma^{\mu}\gamma^5u^{(\chi)}_\mu\chi~,
\label{eq:axnuc}
\eeq
with the axial iso-scalar coupling $g_A = \Delta_u^p - \Delta_d^p \simeq 1.246$~\cite{ParticleDataGroup:2020ssz,Aoki:2021kgd}, 
the iso-vector coupling $g_0 = \Delta_u^p+\Delta_d^p \simeq  0.339$~\cite{Aoki:2021kgd},
and
\beq
\begin{aligned}
u_\mu^{(\chi)} &= -\frac{1}{2f}\left[(c_\alpha+\tilde{\varepsilon}\,s_\alpha)\,\del_\mu \bar{a} + (s_\alpha -\tilde{\varepsilon}\,c_\alpha)\,\del_\mu\bar{\pi}\right]~, \ \
\\
u^{(0)}_\mu &= \frac{C_0}{2f}\left[(c_\alpha+\tilde{\varepsilon}\,s_\alpha)\,\del_\mu \bar{a} + (s_\alpha-\tilde{\varepsilon}\,c_\alpha)\,\del_\mu\bar{\pi}\right]~,
\\
u^{(1)}_\mu &= \frac{1}{f_\pi}\left[-(s_\alpha-\tilde{\varepsilon}\,c_\alpha )\,\del_\mu\bar{a} +(c_\alpha+\tilde{\varepsilon}\, s_\alpha)\,\del_\mu\bar{\pi}\right]~,
\end{aligned}
\label{eq:ALPpicoupnucleon}
\eeq
where $C_0 = c_{uu}+c_{dd}+2c_{GG}$, and $\tilde{\varepsilon}=\varepsilon\;\Theta(m_\pi^2 -m_a^2)$.

The couplings in Eq.~\eqref{eq:ALPpicoupnucleon} imply that DM-nucleon interactions are mediated by a combination of axion and neutral pion exchange. 
The leading Feynman diagrams for spin-dependent $N+\chi \to N+\chi$ scattering are given in Fig.~\ref{fig:feyndiag}(a), and
the corresponding matrix element is
\beq
\mathcal{M}&=&[\bar{u}_{N^\prime}i\gamma^5u_N]\,[\bar{u}_{\chi^\prime}i\gamma^5u_\chi]\,C_a^N \ ,
\label{eq:matelPScurrents}
\eeq
where 
\beq
C_a^N = \frac{m_\chi m_N}{f^2}\,\frac{2\,c_{GG}}{q^2-m_a^2}\,{G}_a^N(q^2) \ ,
\label{eq:CaN}
\eeq
with the form factor
\beq
&&\hspace{-1cm}4c_{GG}\,{G}_a^N(q^2) 
\nnmb\\
&=&\Bigg(
g_0\,C_0\;\frac{q^2-c_\alpha^2m_\pi^2-s_\alpha^2m_a^2-2\tilde{\varepsilon}s_\alpha c_\alpha\,(m_\pi^2-m_a^2)}{q^2-m_\pi^2}
\nnmb\\
&&+ \sigma^3_N\,g_A\lrf{2f}{f_\pi}\,\left[s_\alpha c_\alpha -\tilde{\varepsilon}(c_\alpha^2-s_\alpha^2)\right]\,\frac{m_\pi^2-m_a^2}{q^2-m_\pi^2}\Bigg)
\nnmb\\
&\ \to \ & 
g_0C_0 - \sigma^3_N\,g_A\lrf{2f}{f_\pi}\varepsilon\,\frac{m_\pi^2}{q^2-m_\pi^2}
\label{eq:axff} \ ,
\eeq
where $\sigma^3_N=\pm 1$ for $N=p,\,n$ and the last expression applies for $|m_a^2-m_\pi^2| \gg \varepsilon\,m_\pi^2$. For $c_{uu},\,c_{dd}\to 0$, this form factor matches the leading CP-odd gluonic nucleon form factor presented in Refs.~\cite{Bishara:2017pfq,DelNobile:2021wmp} and given in Eq.~\eqref{eq:psglu} below. 

In the non-relativistic limit, this matrix element leads to spin-dependent scattering on nuclei with a per-nucleon scattering cross section of
\beq
\frac{d\sigma_{\rm SD}}{dQ^2} &=&
\frac{1}{16\pi}\,\frac{Q^4}{Q_{\rm max}^2(m_\chi+m_N)^2}\,|C_{a}^N|^2 \ ,
\label{eq:xsSD-alp}
\eeq
where $Q^2 = |q^2|$ is the 3-momentum transfer,
and $Q \leq Q_\text{max} = 2\,\mu_{N\chi}v_\chi$ for incoming non-relativistic DM velocity $v_\chi$. A key feature of this result is that the total cross section is suppressed by four powers of $v_\chi$ after integrating up to $Q_{\rm max}^2$.

\subsubsection{ALP Spin-Independent}

The exchange of the radial mode $r$ leads to nuclear scattering {\em in}dependent of spin and velocity, via the $t$-channel Feynman diagram in Fig.~\ref{fig:feyndiag}(b).
Since the radial mode is heavy, it can be safely integrated out based on the interactions of Eq.~\eqref{eq:radialglu} to yield an effective operator of the form $(9\alpha_s/8\pi)\,C_G\,\bar{\chi}\chi\,GG$ with 
\beq
C_G = \frac{4\,c_{GG}}{27}\,\frac{m_\chi}{f^2m_r^2} \ .
\label{eq:radff}
\eeq
Matching this gluon operator to nucleons yields an effective SI-nucleon dark matter cross section of
\beq
\frac{d\sigma_{\rm SI}}{dQ^2} &=& 
\frac{1}{Q_{\rm max}^2}\frac{\mu_{N\chi}^2}{\pi}\,|C^N_r|^2
\label{eq:radialsi} \ ,
\eeq
where $\mu_{N\chi}$ is the nucleon-DM reduced mass, $C^N_r = C_G\,m_Nf_G^N$, and $m_Nf_G^N  \simeq 0.85\,\gev$ is the CP-even gluon nucleon form factor~\cite{Bishara:2017pfq,DelNobile:2021wmp}.
Integrating over $Q^2$ gives 
\beq
\sigma_{\rm SI}
&\simeq& (2.0\times 10^{-46}\,\text{cm}^2) \ \times \
\nnmb\\
&&~~c_{GG}^2\lrf{\mu_{N\chi}\,m_\chi}{\gev^2}^2\lrf{f}{m_r}^4\lrf{100\,\gev}{f}^8 \ ,
\nnmb
\label{eq:radialSIxs}
\eeq
Note that we have implicitly normalized the radial mode mass $m_r$ to $f$, a benchmark we will use throughout this paper.

\subsection{\thdma Model Results}

As in the ALP scenario, the 2HDM+$a$ scenario gives rise to both spin-dependent and spin-independent nuclear scattering. The former arises from exchange of the pseudoscalars and the latter via quantum loops, as depicted in Fig.~\ref{fig:feyndiag}(a),(c).
A detailed treatment follows.

\subsubsection{\thdma Spin-Dependent}

The SD per-nucleon cross section with $t$-channel mediation by $a$ and $A$ as in Fig.~\ref{fig:feyndiag}(a) may be written as
\beq
\frac{d\sigma_{SD}}{dQ^2} = \frac{1}{16\pi}\frac{Q^4}{Q_{\rm max}^2(m_\chi+m_N)^2}\,|C_{\rm PS}^N|^2
\label{eq:xsSD-2HDMa}
\eeq
with
\beq
C^{N}_{\rm PS} = \sum_{q=u,d,s}C_q\,(G_q^N+G_G^N)+\sum_{q=c,b,t}C_q\,G_G^N \ ,
\eeq
where
\beq
C_q = \frac{m_N}{v}  \bigg(\frac{\xi^\chi_a \xi_a^q}{m_a^2-q^2} + \frac{\xi^\chi_A \xi_A^q}{m_A^2-q^2} \bigg) \ ,
\label{eq:psff}
\eeq
and 
\beq
G_q^N(q^2) &=& \Delta_q^{N} - q^2\left(
\frac{a_{q,\pi}^{N}}{q^2-m_\pi^2} +
\frac{a_{q,\eta}^{N}}{q^2-m_\eta^2}
\right)
\label{eq:psq}\\
G_G^N(q^2) &=& (-1)\!\!\sum_{q=u,d,s}\!\frac{\bar{m}}{m_{q}}G_{q}^N(q^2) \ .
\label{eq:psglu}
\eeq
Here, $\bar{m} = (1/m_u+1/m_d+1/m_s)^{-1}$, $m_\pi = 135\,\mev$, $m_\eta = 548\,\gev$, and the remaining coefficients can be expressed in terms of the quark spin fractions $\Delta_q^N$. We use the values from the recent average of $2+1+1$ lattice results of Ref.~\cite{Aoki:2021kgd},
\beq
\Delta_u^p = 0.777 \ , \qquad
\Delta_d^p = -0.483 \ , \qquad
\Delta_s^p = -0.053 \ , \qquad
\eeq
with $\Delta_u^n = \Delta_d^p$, $\Delta_d^n = \Delta_u^p$, $\Delta_s^n = \Delta_s^p$, and $g_A = \Delta_u^p-\Delta_d^p$. The pole coefficients are
are~\cite{Bishara:2017pfq,DelNobile:2021wmp} 
\beq
a_{u,\pi}^p = -a_{d,\pi}^p = \frac{1}{2}g_A \ ,~~a_{s,\pi}^p = 0 \ ,
\nnmb
\eeq
with $a_{u,\pi}^n = a_{d,\pi}^p$, $a_{d,\pi}^n = a_{u,\pi}^p$, $a_{s,\pi}^n = a_{s,\pi}^p$, and
\beq
a_{u,\eta}^{p,n} = a_{d,\eta}^{p,n} = -\frac{1}{2}a_{s,\eta}^{p,n} = \frac{1}{6}(\Delta_u^{p}+\Delta_d^{p}-2\Delta_s^{p}) \ .
\nnmb
\eeq
We note that these quantities have moderate uncertainties, but their effect is significantly less than other uncertainties related to neutron star capture.


\subsubsection{\thdma Spin-Independent}
\label{subsubsec:SIloop}

In the minimal \thdma model presented above there is no tree-level contribution to SI nucleon scattering. However, a velocity-independent SI interaction is generated at loop level as illustrated in Fig.~\ref{fig:feyndiag}. These contributions were studied in Refs.~\cite{Ipek:2014gua,Arcadi:2017wqi,Bell:2018zra} and developed further in Refs.~\cite{Abe:2018emu,Ertas:2019dew}.

The SI nucleon cross section can be written as
\beq
    \frac{d\sigma_\text{SI}}{dQ^2} = 
    \frac{1}{Q_{\rm max}^2}
    \frac{\mu^2_{N\chi}}{\pi} \left| C^N_{\rm loop}\right|^2
\eeq
where
\beq
    C^N_{\rm loop} \simeq C_G\,m_N f^N_{G} + \sum_{q=u,d,s} C_q\,m_N f^N_{q} \ ,
 \nnmb
    \label{eq:2hdmasi}
\eeq
with the recent lattice determinations~\cite{Alexandrou:2019brg}
\beq
f_u^p =  0.017 \ ,\qquad
f_d^p =  0.026 \ ,\qquad
f_s^p =  0.049 \ ,\qquad\\
f_u^n =  0.013 \ ,\qquad
f_d^n =  0.033 \ ,\qquad
f_s^n =  0.049 \ ,\qquad
\nnmb
\eeq
and $f_G^N = 1-\sum_{q}f_q^N \simeq 0.91$.

Explicit expressions for the loop-induced $C_q$ and $C_G$ coefficients in the \thdma are collected in Ref.~\cite{Abe:2018emu}. For the model parameters considered in this work, the dominant loops for SI scattering correspond to the lower two diagrams in Fig.~\ref{fig:feyndiag}. The $C_q$ coefficients in Eq.~\eqref{eq:2hdmasi} are generated mainly by triangle diagrams of the form shown in Fig.~\ref{fig:feyndiag}~(c). The $C_G$ coefficient comes primarily from triangle and box loops connecting to heavy ($c,b,t$) quarks. We consider $\tan\beta=3$, but at larger values a contribution from two-loop diagrams involving gluons can come to dominate due to a $\tan^4\!\beta$ enhancement in loops involving bottom quarks~\cite{Abe:2018emu}. Higher twist operators are also considered in Ref.~\cite{Abe:2018emu}, but we find that their numerical contributions here are very small.

Let us note that loop diagrams in the ALP benchmark do not contribute to the SI cross section in this model to as much of an extent as in the \thdma. There is no Higgs-ALP coupling generated at leading order~\cite{Bauer:2021wjo}, and this eliminates the triangle diagram. For the box diagrams, the crucial difference is that the ALP couplings of Eqs.~(\ref{eq:alpglu},\ref{eq:alpferm}) involve derivative operators rather than a true pseudoscalar-fermion interaction. These are equivalent on-shell, but in the box loops the derivative couplings add powers of loop momentum and make the result UV-dependent. The leading contribution to SI scattering in the ALP benchmark is therefore from the radial mode rather than loops.

\begin{figure}
    \centering
 \includegraphics[width=.49\textwidth]{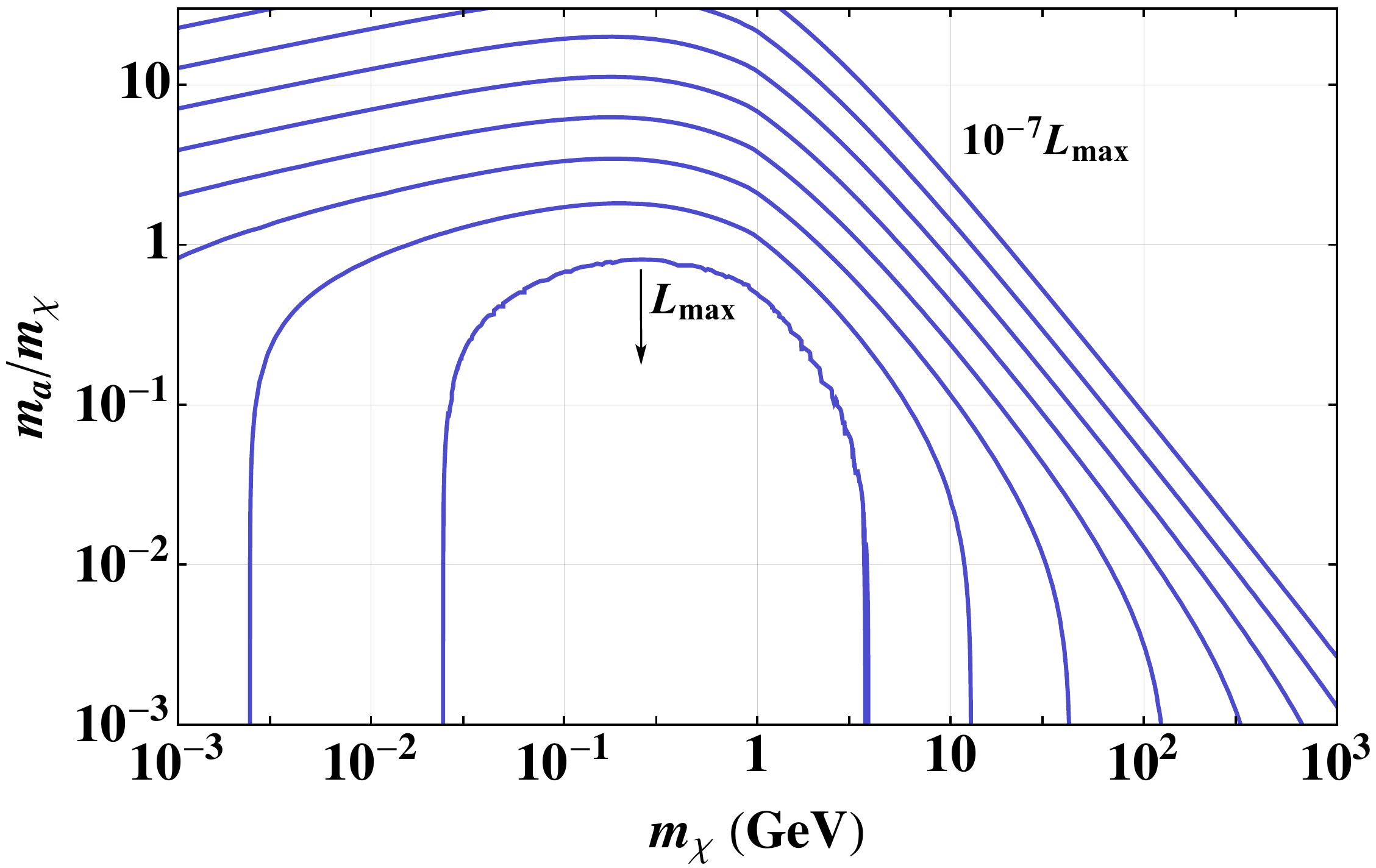} 
 \caption{\raggedright{\small Contours of constant NS luminosity, logarithmically equispaced, for effective couplings $y_\chi^2 y_n^2 = 10^{-14} = 1$;
 here $L_{\rm max}$ is the maximum NS luminosity that may be imparted by kinetic heating due to incident dark matter flux.
 Equivalently, these are contours of $y_\chi^2 y_n^2$ corresponding to NS luminosity $L_{\rm max}$, with values of $y_\chi^2 y_n^2$ that increase as we go from inner regions of the plot to the outer.}
 }
    \label{fig:conts-caprate}
\end{figure}

\section{Neutron star heating}
\label{sec:NScapture}

We turn next to study the heating of neutron stars by the infall and capture of local $\chi$ dark matter based on the two mediator benchmark scenarios introduced above. First we review general features of dark kinetic heating of neutron stars, and then we specialize to the specific aspects of the capture and heating when DM connects to the SM through a light pseudoscalar mediator.

\subsection{Review of Dark Kinetic Heating}

For our treatment of DM capture in neutron stars we will consider the following representative NS configuration:
\beq
M_{\rm NS} = 1.5 \ M_\odot,~~R_{\rm NS} = 12.85 \ {\rm km} \ ,
\label{eq:BMNS}
\eeq
giving an escape speed at the surface
\beq
v_{\rm esc} = \sqrt{\frac{2 G \MNS}{\RNS}} \simeq 0.59 \ . 
\eeq
This configuration is obtained from a Quark Meson Coupling~(QMC) equation of state~(EOS) of matter near nuclear saturation densities in Ref.~\cite{NSvIR:anzuiniBell2021improved}, where it is also shown that near-identical results are obtained for the Brussels-Montreal EOS ``BSk24".
The self-interactions of the nuclear medium in the NS core effectively modify the neutron mass, $m_n \to \mneff$, and in our work we set
\beq
\mneff = 719.6 \ {\rm MeV} \ ,
\eeq
obtained by volume-averaging the NS radius-dependent effective $m_n$ shown in Ref.~\cite{NSvIR:anzuiniBell2021improved}.
We will use this representative NS configuration to illustrate our main effects and to present the central values of our parametric reaches while accounting for uncertainties in the (yet unknown) mass and radius of candidate NSs.

To estimate the sensitivity of NS capture in the coupling-mass space of the underlying DM theory, we first compute the fraction of the DM flux incident on the NS that is gravitationally captured.
For local DM density $\rho_\chi$ and average DM-NS relative speed $v_{\rm rel}$ (which at the solar position are respectively taken as $0.4$~GeV/cm$^3$ and 350 km/s~\cite{vcircMW}), the DM mass capture rate is given by~\cite{Goldman:1989nd} 
\beq
\dot M = \mdm C_{n\chi} = \rho_\chi v_{\rm rel} \times \pi b_{\rm max}^2  \times p_v \times p_\sigma \ ,
\label{eq:masscaprate}
\eeq
where $b_{\rm max} = \RNS (1+z) (v_{\rm esc}/v_{\rm rel})$ is the maximum impact parameter of DM traversing the NS, with $1+z = (1-v_{\rm esc}^2)^{-1/2}$ a blueshift factor magnifying the NS radius to a distant observer, and $p_v$ is the probability that a scattered DM particle loses sufficient energy to be captured; in Appendix~\ref{app:lightmedNS} we show that  for pseudoscalar mediation $p_v = 1$ is an excellent approximation throughout the parameter space we consider.
The probability that incident DM is scattered is given by
\bea
\nn p_\sigma &=& 1 - e^{-\tau} \simeq \tau \\
 &=& \sigma_{n \chi}/\sigma_{\rm cap} \ ,
 \label{eq:psigma}
\eea
where, for optical depth $\tau$, the approximate equality in the first line holds in the optically thin limit considered here.
The ``capture cross section" above which $\tau > 1$ in the NS core is:
\begin{align}
\displaystyle
\sigma_{\rm cap} =
 \begin{cases}
	\sigmageom (\mneff/\mdm)\
	&, \ \ \mdm < \mneff \ , 
	\\
	\sigmageom \
	&, \ \ \mneff \leq \mdm \leq \text{PeV} \ , 
	\\
	\sigmageom (\mdm/{\rm PeV})\
	&, \ \ \mdm > {\rm PeV} \ ,
 \end{cases}
 \label{eq:sigmathreshold}
\end{align}
where the NS geometric cross section $\sigmageom = \pi (\mneff/\MNS) \RNS^2 \simeq 2.2 \times 10^{-45}\,{\rm cm}^2$.
The dependence on $\mdm$ in Eq.~\eqref{eq:sigmathreshold} is understood by considering the typical DM recoil energy in the neutron rest frame:
\begin{align}
\Er &\simeq \frac{\mneff \mdm^2 \gamma^2 v^2_{\rm esc}}{(\mneff^2+\mdm^2+2\gamma \mneff\mdm)} \ ,
\label{eq:Erec}
\end{align}
with $\gamma = 1+z$.
For $\mdm \!<\!\mneff$, only a fraction $\simeq 3 \Delta p/ p_F$
		of degenerate neutrons close enough to their Fermi surface receive the typical momentum transfer $\Delta p = \sqrt{2\mneff \Er}$ to scatter\footnote{As noted in Ref.~\cite{NSvIR:clumps2021}, for $\mdm \lsim 35$~MeV nucleons in the NS might not scatter elastically due to a superfluidity energy gap, leaving DM to capture via collective excitations instead. A detailed treatment of this effect is beyond the scope of this work, however, we note that in the DM mass range where it applies, current limits on our mediator already outdo the future sensitivity of NS heating: see Fig.~\ref{fig:lims-bauerBMcGG-masratios}. 
		For a study of collective effects in NSs impacting light DM capture in certain models, see Ref.~\cite{collectiveDeRoccoLasenby:2022rze}.} to above the Fermi momentum $p_F \simeq 0.4~\text{GeV}$.
	This so-called Pauli blocking gives	$\sigma_{\rm cap} \propto \Er^{-1/2} \propto \mdm^{-1}$.
 For $\mneff\!\leq\! \mdm \!\leq\! 10^6~\!\text{GeV}$, a single scatter suffices for capture: $\Er \simeq \mneff v_{\rm esc}^2 \gamma^2 >$ KE$_{\rm halo}$, the DM halo kinetic energy. 
For $\mdm > \text{PeV}$, multiple scatters are required for capture, so that approximately $\sigma_{\rm cap}\propto {\rm KE}_{\rm halo}/\Er \propto \mdm$. 
  
Under equilibrium, the kinetic power of the infalling dark matter (the heating rate) equals the rate at which photons are emitted from the NS surface (the cooling rate, dominated by such photon emission for NSs older than $\sim$Myr~\cite{coolingminimal:Page:2004fy,cooling:Yakovlev:2004iq}). 
Assuming blackbody radiation, the NS luminosity corresponding to a temperature $T$ (in the NS frame) is then  
  \beq
  L = z \dot M = 4 \pi \RNS^2 T^4 \ .
\eeq
This luminosity attains a maximum value $L_{\rm max}$ for unit capture probabilities $p_\sigma$ and $p_v$.

Applying these general results to
our representative NS configurationwith parameters as in Eq.~\eqref{eq:BMNS},
we find the maximum luminosity
 \beq
L_{\rm max} = 7.6 \times 10^{24}~{\rm GeV/s} \ .
\label{eq:Lmax}
 \eeq
This corresponds to a NS temperature seen by a distant observer $\tilde T = T/(1+z)$ of
\beq
\tilde T = 1400~{\rm K} \ .
\eeq
Such a NS temperature is measurable within reasonable integration times at current and imminent infrared telescope missions~\cite{NSvIR:Baryakhtar:DKHNS,NSvIR:Raj:DKHNSOps}, in particular at the recently launched James Webb Space Telescope (JWST)\footnote{See Ref.~\cite{NSvIR:IISc2022} for an astronomical study on observing NS heating at JWST.}~\cite{JWST:Gardner:2006ky}, and the forthcoming Extremely Large Telescope~\cite{ELT:neichel2018overview} and
Thirty Meter Telescope~\cite{TMT:2015pvw}.

Possible NS-reheating mechanisms that may compete with DM kinetic heating are:
(a) the accretion of interstellar material, likely to be deflected along the NS's magnetic field lines such that only a small polar region is heated, distinguishable from all-surface thermal emission~\cite{Treves:1999ne};
(b) rotochemical heating in the case of NSs with very small (sub-7 ms) initial spin period for certain nucleon pairing models~\cite{NSvIR:Hamaguchi:Rotochemical}.

While we focus on the heating of neutron stars from the transfer of kinetic energy by captured dark matter in this work, which applies to any type of dark matter candidate, we note that dark matter that annihilates within a NS could raise its brightness further~\cite{Kouvaris:2007ay,deLavallaz:2010wp} and thereby improve telescope sensitivities~\cite{NSvIR:Baryakhtar:DKHNS,NSvIR:Raj:DKHNSOps,NSvIR:IISc2022}. This effect depends non-trivially on whether or not the DM thermalizes with the NS over the stellar lifetime~\cite{Bertoni:2013bsa,NSvIR:GaraniGuptaRaj:Thermalizn}, and we defer its study to a future analysis.

\subsection{Capture via (Pseudo)scalar Mediation}

In Sec.~\ref{sec:direct} we computed per-nucleon scattering cross sections relevant for direct detection experiments. To apply these to DM capture in neutron stars, where the incoming DM and target nucleons are more energetic and can both be semi-relativistic~\cite{NSvIR:Riverside:LeptophilicLong,NSvIR:Bell2020improved,NSvIR:anzuiniBell2021improved}, we generalize our previous results in two ways.
First, we extend the nucleon matrix element coefficients $C_i^N$ of Eqs.~(\ref{eq:axff},\ref{eq:radff},\ref{eq:psff},\ref{eq:2hdmasi})by multiplying them by a dipole form factor~\cite{Bishara:2017pfq,DelNobile:2021wmp},
\beq
C_i^N(t) \ \to \  C_i^N(t)\times\frac{1}{(1-t/M_D^2)^2}
\eeq
with $t=q^2=-Q^2$ and $M_D \simeq 1\,\gev$. This form factor accounts for the inner structure of the nucleon probed at higher momenta and corresponds to expanding beyond the leading non-trivial order in an expansion in $t/m_N^2$~\cite{Bishara:2017pfq,DelNobile:2021wmp}.
Our second modification is to use fully relativistic expressions for the cross sections. In the notation of Sec.~\ref{sec:direct}, the summed, squared, and averaged matrix elements are
\beq
|\mathcal{M}|^2 = \left\{
\begin{array}{lcl}
|C_i^N|^2\,(4m_\chi^2-t)\,(4m_N^2-t)&;&\text{scalar}\\
&&\\
|C_i^N|^2\,t^2&;&\text{pseudoscalar}
\end{array}\right.
\eeq
where $t=q^2=-Q^2$ is the usual Mandelstam variable. 

Consider now the scattering of DM with neutrons mediated by $t$-channel pseudoscalar exchange. The dependence of the cross sections on momentum transfer makes this effectively unobservable in direct detection, but we demonstrate below that these channels can be dominant for DM capture in neutron stars. To illustrate how pseudoscalar mediation can be enhanced in NS capture with a focus on the impact of the mediator mass, we use a simplified pseudoscalar model with $C_{\rm PS,a}^N \to y_\chi y_n/(m_a^2 -t)$ for effective DM and neutron couplings $y_\chi$ and $y_n$.
In Fig.~\ref{fig:conts-caprate} we show contours of constant NS luminosity, corresponding one-to-one to contours of the  capture probability $p_\sigma$, in the plane of $\mdm$ versus $m_a/\mdm$ for the fixed coupling product $y_\chi^2 y_n^2 = 10^{-14}$. Here, the entire region below the contour depicting $L_{\rm max}$ corresponds to $p_\sigma = 1$. The scaling behaviour of these contours can be understood as follows.
For $\mdm \ll m_n$, the typical momentum transfer is $Q \simeq \mdm v_{\rm esc} (1+z) \simeq \mdm$, so we have
\beq
\sigma_{n\chi}|_{\mdm \ll m_n} &\propto& \frac{\mdm^4}{(\mdm^2+m_a^2)^2}~.
\eeq
Combining this with the fact that for $\mdm < m_n$ Pauli-blocking sets $p_\sigma \propto \mdm^{-1}$ (Eqs.~\eqref{eq:psigma} and \eqref{eq:sigmathreshold}), we obtain
\beq
L|_{\mdm \ll m_n} \propto p_\sigma \propto \begin{cases}
1/\mdm,~~~~m_a \ll \mdm,\\
\mdm^3/m_a^4,~m_a \gg \mdm \ .
\end{cases}
\eeq
This is indeed the scaling we observe in the $m_a \ll \mdm$ and $m_a \gg \mdm$ limits in Fig.~\ref{fig:conts-caprate}.

For $\mdm \gg m_n$, the typical momentum transfer is $Q \simeq m_n v_{\rm esc} (1+z) \simeq m_n$, and hence
\beq
\sigma_{n\chi}|_{\mdm \gg m_n} &\propto& \frac{1}{\mdm^2(m_n^2+m_a^2)^2} \ .
\eeq
In this mass range there is no Pauli-blocking; hence we find these scalings in Fig.~\ref{fig:conts-caprate}:
\beq
L|_{\mdm \gg m_n} \propto p_\sigma \propto \begin{cases}
1/\mdm^2,~~~~m_a \ll m_n,\\
1/\mdm^2m_a^4,~m_a \gg m_n~.
\end{cases}
\eeq
We will find the above features reflected in the NS heating sensitivities we estimate below.

\section{Results}
\label{sec:results}
Having investigated our benchmark pseudoscalar-mediated DM models as well as the general features of DM kinetic heating of neutron stars, we now present our specific results in detail.

\subsection{ALP Scenario}

In Figure~\ref{fig:lims-bauerBMcGG-masratios} we show in red and purple bands the sensitivity of NS heating to the dimensionless parameter $c_{GG}\,m_n/f$ characterizing the ALP-nucleon coupling strength as a function of the DM mass $m_\chi$.
The solid curves are the sensitivities to spin-dependent scattering on NS nucleons via ALP exchange, whereas the dashed curves are to spin-independent scattering via exchange of the radial mode $r$.
The left panel corresponds to fixing the mediator mass to $m_a = \mdm/10$, representative of a ``light mediator" regime, whereas the right panel depicts $m_a = \mdm$ to represent an ``intermediate mediator mass" regime.

The bands correspond to the reach over the model parameter space from the measurement of a NS with luminosity $L_{\rm max}$ given in Eq.~\eqref{eq:Lmax}, where the upper~(lower) boundary is for a NS mass of $M_{\rm NS} = 1.5\,M_\odot$~($2.16\,M_\odot$).
In general, the precise mass of the NS under observation is not known, with heavier NSs yielding a greater luminosity for a given capture cross section.
This is the largest astrophysical uncertainty in the capture rate, spanning an order of magnitude and exceeding those coming from other effects such as the unknown EoS of NS matter, the superfluid nature of NS nucleons, etc.~\cite{snowmassWP:Berti:2022rwn}. 
We estimate that the capture cross section varies by only a factor of 1.6 across NS mass-radius configurations across the mass range $[1.5,\, 2.16]\,M_\odot$~\cite{NSvIR:anzuiniBell2021improved}. 
Therefore we may infer that, for an observed NS luminosity, there is an uncertainty of $\Oc(10)$ in the minimum ALP-mediated DM cross section deduced from it.
In practice we fix this uncertainty to $\sigma_{\rm uncert}$ = 10, and as the ALP exchange ($r$ exchange) cross section $\propto f^{-4}$ ( $\propto f^{-8}$), the bands spread vertically over factors of $\sigma_{\rm uncert}^{1/4}$ ($\sigma_{\rm uncert}^{1/8}$).

The solid curves are seen to change slope at two different DM masses: near $\mdm =$ 20 MeV in a sharp manner, and near $\mdm =$ GeV more gently. 
The first feature is due to cancellations between the iso-scalar and iso-vector terms in the form factor in Eq.~\eqref{eq:axff} for small momentum transfers.
As we raise $\mdm$ above $m_\pi$ (while keeping $\mdm <  \bar m_n$), Eq.~\eqref{eq:axff} is dominated by the iso-scalar term, and we have $\sigma_{\rm n \chi} \propto |\mathcal{M}|^2/\bar{m}_n^2  \propto f^{-4} \mdm^2$. 
In this regime the NS capture cross section  $\sigma_{\rm cap}\propto \mdm^{-1}$ (Eq.~\eqref{eq:sigmathreshold}), thus the constraint on $1/f \propto \mdm^{-3/4}$.
As we transition to $\mdm$ above $\bar{m}_n$, we have $Q \simeq \bar{m}_n$, thus the spin-averaged $|\mathcal{M}|^2 \propto \mdm^2/f^4/(\bar{m}_n^2 + m_a^2)^2$.
For $m_a = \mdm$, we then have $\sigma_{\rm n\chi} \propto |\mathcal{M}|^2/\mdm^2 \propto f^{-4} \mdm^{-4}$.
As the NS capture cross section here is $\mdm$-independent, the constraint on $1/f \sim \mdm$.
For $m_a = \mdm/10$ the turnover occurs at $\mdm =$ a few $\times \bar{m}_n$ as expected.
We also see that ALP exchange dominates the sensitivity to NS capture over exchange of the heavier $r$-mode, and that the latter has a different slope due to the simple $\mu_{n\chi}^2\mdm^2/f^8$ dependence on the cross section.

Comparing across the two panels, we also see that near $\mdm \simeq 10$~GeV, the constraints on $c_{GG}\,m_n/f$ are about 10 times weaker for $m_a/\mdm = 1$ than for $m_a/\mdm = 1/10$. 
This is because in this region the cross section goes as $f^{-4}m_a^{-4}$.
However, as we go to smaller $\mdm$, and hence smaller $m_a$, the ALP propagator in Eq.~\eqref{eq:matelPScurrents} is dominated by $Q^2$ and the cross section becomes insensitive to $m_a$, resulting in similar limits.
There is no change in the dashed curves across the panels due to the $r$-exchange cross section being independent of $m_a$ (Eq.~\eqref{eq:radialsi}).
The brown region, which also does not change across the panels, is the upper bound from direct detection searches at SuperCDMS~\cite{SuperCDMS:2022kgp}, Xenon-1T~\cite{SDlight:XENON:2019zpr}, DarkSide-50~\cite{DarkSide:2018bpj}, and PandaX-4T~\cite{PandaX-4T:2021bab}.
These limits are obtained from the spin-independent $r$-exchange; the spin-dependent ALP exchange is $Q^4$-suppressed and yields much weaker limits.
The direct detection limits are seen to be generically weaker than the NS capture sensitivity due to their large exclusion cross sections at sub-nuclear DM masses, where these searches are limited by recoil thresholds.

The gray shaded regions in the backgrounds of these plots are excluded by constraints on the ALP mediator from meson decay searches in beam dumps, and from collider searches; see Appendix~\ref{sec:appa} for particulars.
The beam dump searches lose their sensitivity above $m_a$ of $\Oc$(100 MeV)$-\Oc$(GeV) due to kinematic limitation in producing mesons.
Likewise, for $m_a \gsim$ GeV a CMS search for the chromo-magnetic dipole moment of the top quark~\cite{CMS:2019nrx} is limited by backgrounds, yielding an upper bound of $c_{GG}\,m_n/f \lsim 5 \times 10^{-2}$.  
Thus NS heating could emerge the sole probe in this mass regime for $c_{GG}\,m_n/f$ between $\sim 10^{-3}-0.05$, as seen in Fig.~\ref{fig:lims-bauerBMcGG-masratios}.
A notable feature in these limits is the islands of parameter space near $m_a = 100$~MeV unconstrained by beam dumps and colliders.
We see that the upper island ranging between $c_{GG}\,m_n/f \simeq 7 \times 10^{-4} - 3 \times 10^{-2}$ will be entirely covered by NS heating in the light mediator case, and mostly covered in the equal-mass mediator case.
The lower islands around $m_a \simeq 10^{-4}$ may be partially covered by NS heating in the light mediator case; with improved telescope sensitivities to lower NS luminosities, it is possible for some coverage of these islands in both the light and equal-mass mediator scenarios.

\subsection{2HDM+$a$ Scenario}

In Figure~\ref{fig:lims2HDMa} we show in red and purple bands the sensitivity of NS heating to the $a$-$A$ mixing angle $\theta$ as a function of the DM mass $m_\chi$; we cut off the plot at $\theta = \pi/4$ in order to keep the lighter eigenstate $a$ (heavier eigenstate $A$) mostly the interacting field $a_0$ ($A_0$) after electroweak symmetry breaking.
The left (right) panel corresponds to a ``light" (``intermediate") mediator with $m_a/\mdm = 1/3$ ($m_a/\mdm = 1$).
As in the previous sub-section, the bands represent the uncertainty in NS mass-radius configurations, taken to be $\sigma_{\rm uncert} =$ 10.
As the per-nucleon scattering cross section $\sigma_{\rm n\chi} \propto \sin^2 2\theta$ (Eq.~\eqref{eq:xsSD-2HDMa}), the bands spread vertically over a factor of $\sqrt{\sigma_{\rm uncert}}$ for small $\theta$.
Further, as $\sigma_{\rm n\chi} \propto \mdm^{-6}$ for fixed $m_a/\mdm$ ratios, and as the NS capture cross section is $\mdm$-independent in this regime (see Eq.~\eqref{eq:sigmathreshold}), the limits on $\theta \propto \mdm^3$ for small $\theta$, as seen in the figure. 

The brown region in both panels is excluded by the most recent limits from direct detection searches at PandaX-4T~\cite{PandaX-4T:2021bab} and Xenon-1T~\cite{XENON:2018voc}, using the spin-independent loop-induced cross section we derive in Sec.~\ref{subsubsec:SIloop}.
These correspond to exclusion cross sections that range roughly between $4\times10^{-47}$~cm$^2$ -- $4\times10^{-46} \ {\rm cm}^2$.
The loops that result in SI scattering in direct detection result also in SI scattering in NSs, but due to modest $Q$-dependence the NS heating reach (for $\sigma_{\rm cap} \simeq 2 \times 10^{-45}$~cm$^2$) is weaker than the direct detection limits, thus we do not present it here. 
We display contours of constant (spin-independent) $\sigma_{\rm n\chi} = 10^{-47}, 10^{-48}, 10^{-49}$~cm$^2$, depicting the approximate march of progress expected of direct detection experiments of the next two generations until the so-called neutrino floor is reached~\cite{Aalbers:2022dzr}.
The gray regions in the background are excluded by a host of collider and flavor data as detailed in Appendix~\ref{sec:appb}.

Comparing across the two panels, we find that the NS heating sensitivities on $\theta$ are about ten times weaker in the right panel, where $a$ is thrice heavier.
This is just as expected from the scaling $\sigma_{\rm n\chi} \propto \sin^2 2 \theta/m_a^4$ in Eq.~\eqref{eq:xsSD-2HDMa}, implying that the exclusion on $\theta \propto m_a^2$ for small $\theta$.
Our NS heating reach is seen to be stronger than the $B \to K^* \mu \mu$ limits on the mediator, applicable to $m_a \lsim 4$~GeV, for $m_\chi/m_a = 1/3$ in the left panel.
In this panel we also outdo collider limits on the mediator from $\tau \tau \mu \mu$ searches for 12~GeV$\lsim m_\chi \lsim$~25~GeV.
Whether we outdo ATLAS limits from undetected final states depends on the unknown mass of the NS observed: a 1.5 $M_\odot$ NS does not, whereas greater masses yielding greater NS luminosity (depicted by the red band) do.
In the right panel with $m_\chi/m_a = 1$, our reach is stronger than these limits by only a small margin near $m_\chi =$ 10 GeV for a 1.5 $M_\odot$ NS, while outdoing them for up to $m_\chi \simeq$ 16 GeV as we increase the NS mass to 2.16~$M_\odot$.

As for SI direct detection, we observe that these limits are generally weaker than those from flavor and colliders on the mediator, except for $m_\chi (= m_a) \gsim$~62 GeV in the right panel, where there is a gap in collider limits from a combination of kinematic limitations on Higgs decay and lack of analysis on $b b \to A \to \tau \tau$ below $m_{\tau\tau}$ = 90 GeV.
In any case, due to weak dependence on DM mass in the loop-induced SI limits on $\theta$ contrasted with the rapid climb ($\theta \propto m_\chi^3$) in the NS heating reach, we are generally stronger than direct detection for small $m_\chi$ in the mass range displayed, and weaker for higher $m_\chi$.
It is interesting to note that in low mass regions NS heating can reach smaller couplings than future direct DM searches, even after the SI neutrino floor is encountered in the latter.
Thus in these regions ($m_\chi \lsim$~30 GeV for $m_\chi/m_a = 1/3$ and $m_\chi \lsim$~15 GeV for $m_\chi/m_a = 1$) NS heating may emerge a main, if not sole, means to test the 2HDM+$a$ framework.


\section{Conclusions and Outlook}
\label{sec:concs}

In this study we have shown that kinetic heating of neutron stars by dark matter is a leading probe of dark matter candidates that connect to the SM primarily through pseudoscalar mediators. We have done so within the context of two well-motivated, self-consistent pseudoscalar mediator frameworks:  
axion-like particles for sub-10 GeV DM masses, and 
the \thdma model for larger DM masses.
Our results are summarized in Figs.~\ref{fig:lims-bauerBMcGG-masratios} and \ref{fig:lims2HDMa}.
These plots demonstrate a strong complementarity between NS heating tests of these scenarios and direct searches for DM and their associated mediators.

Our work motivates generalizations in a number of new directions:
\begin{itemize}
    \item It would be interesting to explore other pseudoscalar mediator frameworks beyond the specific ALP and 2HDM$+a$ scenarios presented here. In particular, while we have considered a KSVZ-type ALP that couples exclusively to gluons (and DM) at the Peccei-Quinn breaking scale $f$, other ALP scenarios with additional couplings to the SM would invoke further 
    constraints~\cite{Bauer:2021mvw} and modify the effective nucleon coupling and hence the sensitivity to NS heating. This could be realized in the context of DFSZ-type ALPs~\cite{Zhitnitsky:1980tq,Dine:1981rt} that could induce direct couplings to quarks or leptons. In particular, primarily leptonic ALP couplings could lead to DM capture via scattering with the large population of leptons present in neutron stars~\cite{NSvIR:Bell2019:Leptophilic,NSvIR:GaraniGenoliniHambye,NSvIR:GaraniHeeck:Muophilic,NSvIR:Riverside:LeptophilicShort,NSvIR:Riverside:LeptophilicLong, NSvIR:Bell:ImprovedLepton}.

    \item The motivation of this work is to identify and investigate self-consistent scenarios unfriendly to direct detection but susceptible to NS capture. 
    Pseudoscalar mediation is one such important possibility. However, there are other ones explored in the literature: models of inelastic DM (thermal Higgsinos~\cite{NSvIR:Baryakhtar:DKHNS,NSvIR:Pasta} and other electroweak multiplets~\cite{Hamaguchi:2019oev}), and muon-philic DM in a gauged $L_\mu - L_\tau$ model~\cite{NSvIR:GaraniHeeck:Muophilic}.
    Yet more possibilities include DM in the keV-GeV mass range with small scattering cross sections,
    DM with large nuclear cross sections shielded by the rock overburden, 
    composite DM with super-Planckian masses yielding too small a flux for direct searches~\cite{Bramante:2018qbc,Bramante:2018tos,Bramante:2019yss,DEAPCollaboration:2021raj,SnowmassHeavy:Carney:2022gse},
    models that lead to appreciable clustering of DM so that the Earth encounters DM clumps too infrequently~\cite{NSvIR:clumps2021}, and so on.
    We leave these avenues of exploration to future
    authors.
    \item Certain model-specific assumptions here may be relaxed, leading to a richer phenomenology that we leave to future work. For instance, we chose the maximum mediator mass to be the DM mass in order to avoid invisible decay modes, however a recasting of collider and flavor constraints incorporating such decays would be an interesting exercise. 
    Similarly, whilst we had taken DM to be asymmetric, turning on (even tiny) annihilation rates could not only increase the resultant NS luminosity (see, e.g., Ref.~\cite{Kouvaris:2007ay}), but also invoke limits from indirect detection.
    An interesting computation in this case would be of the timescale for DM thermalization with the NS.
    In particular, in regions leading to NS capture via SI scattering via radial mode mediation (regions above the dashed line in Fig.~\ref{fig:lims-bauerBMcGG-masratios}), repeated scatters with NS nucleons would slow down DM particles and the $Q^4$-dependent ALP-mediated SD cross section may get smaller than the SI cross section, leading to the intriguing possibility that thermalization is driven initially by SD scattering but later by SI scattering.
    \item We have assumed here that either of our scenarios makes up the entire DM population.  
      Were it instead to constitute a fraction, the limits on the direct detection cross section would weaken proportionally, but this does not necessarily apply to NS heating. 
      Rather, the DM-heated NS' luminosity would be proportionally smaller, and hence a detailed astronomical treatment such as in Refs.~\cite{NSvIR:Baryakhtar:DKHNS,NSvIR:Raj:DKHNSOps,NSvIR:IISc2022} must be carried out to determine the most optimal filter, integration times, and so forth.
    
\end{itemize}

As with the overheating of neutron stars in complete models of exotic baryons~\cite{McKeen:2020oyr,McKeen:2021jbh,Goldman:2022brt},
their overheating by the capture of dark matter in complete models of the mediator
provides for upcoming astronomical missions the added motivation of directly probing fundamental physics. 

\section*{Acknowledgments}

This work is supported in part by the Natural Sciences and Engineering Research Council (NSERC) of Canada.
TRIUMF receives federal funding via a contribution agreement with the National Research
Council (NRC) of Canada.

\newpage 

\appendix

\section{Limits on the ALP Model\label{sec:appa}}

\begin{figure}
    \centering
 \includegraphics[width=.49\textwidth]{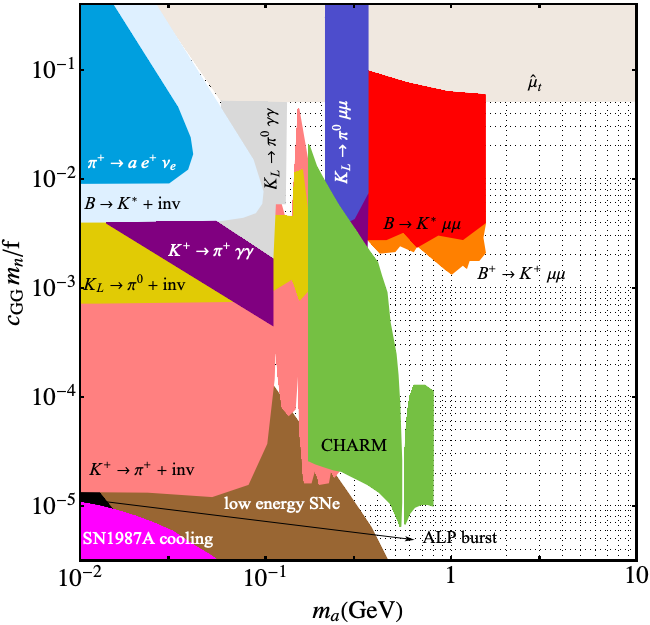} 
 \caption{\raggedright{\small Limits on the ALP mediator in the plane of the effective ALP-nucleon coupling versus the ALP mass, from beam dump, flavor, collider, and astrophysical measurements.
  See text for further details.}
    }
    \label{fig:lims-med-ALP}
\end{figure}

\begin{figure}
    \centering
 \includegraphics[width=.49\textwidth]{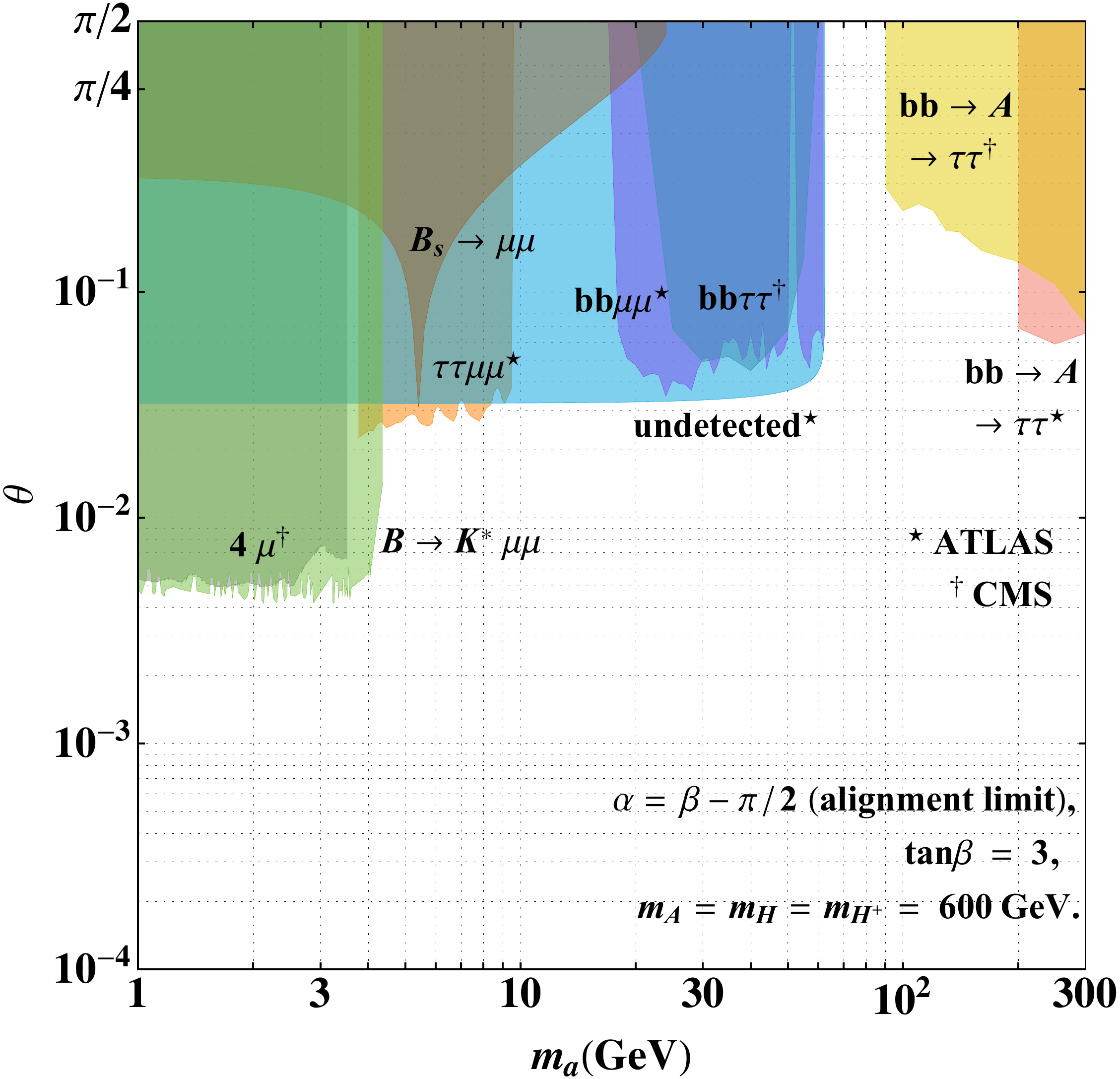} 
 \caption{\raggedright{\small Derived limits on the \thdma model in terms of the light pseudoscalar mass $m_a$ and the pseudoscalar mixing angle $|\theta|$ from flavor tests, Higgs measurements, and collider searches. The other \thdma model parameters are kept fixed at the benchmark values discussed in the text with Type-II couplings to fermions, $\tan\beta = 3$, and $m_H=m_A=m_{H^+}=600\,\gev$.}
    }
    \label{fig:limsmed-2hdma}
\end{figure}

In this appendix we elaborate on the various limits on the ALP mediator as shown in Fig.~\ref{fig:lims-med-ALP}.

{\noindent \bf \em Collider and Beam Dump Limits} 

Many of the limits we display were described in Ref.~\cite{Bauer:2021mvw} in the space of $c_{GG}/f$ vs $m_a$; we reproduce them here in the $c_{GG}\,m_n/f$ vs $m_a$ plane by making the substitution $c_{GG} \to T_2(r) = 1/2$. 
The key phenomenological divider here is the QCD scale $\sim m_\pi$. 
A non-zero $c_{GG}$ induces sizeable ALP couplings to the photon via the QCD anomaly, $c_{\gamma \gamma} \simeq -1.92 c_{GG}$~\cite{Bauer:2020jbp}, so that for $m_a < m_\pi$ the diphoton mode $a \to \gamma \gamma$ dominates the ALP decay width.
For $m_a > m_\pi$ the branching to hadrons dominates, but decays to muons $a \to \ell^+ \ell^-$ at the sub-percent level turn out to be important.

For $m_a < m_\pi$ the strongest terrestrial constraints come from searches for $K^+ \to \pi^+ + X_{\rm inv}$ at NA62~\cite{NA62:2021zjw}, with $X_{\rm inv}$ an invisible state.
In this range of $m_a$ the $a \to \gamma \gamma$ decay length exceeds the 10 m detector size and is thus unseen~\cite{Bauer:2021mvw}.
Limits from $K_L \to \pi^0 X_{\rm inv}$ at KOTO~\cite{KOTO:2018dsc} are weaker due to suppressed CP-conserving couplings, and even weaker limits are set by $B \to K^* X_{\rm inv}$ at Belle~\cite{Belle:2017oht} and $\pi^+ \to a e^+ \nu_e$ at PIENU~\cite{PIENU:2021clt}.
For somewhat large $m_a$ and $c_{GG}$ the $a \to \gamma \gamma$ mode becomes visible, and limits are obtained from $K^+ \to \pi^+ \gamma \gamma$ and $K_L \to \pi^0 \gamma \gamma$ at E949, NA48, NA62, and KTeV~\cite{E949:2005qiy,NA48:2002xke,NA62:2014ybm,KTeV:2008nqz}.

For $2m_\mu < m_a < m_B$ limits from $B^+ \to K^+ a (\mu\mu)$ and $B^0 \to K^* a (\mu\mu)$ at LHCb~\cite{LHCb:2015nkv,LHCb:2016awg} apply, and for $m_a \gsim$~GeV the strongest limits come from those placed by CMS~\cite{CMS:2019nrx} on the chromo-magnetic dipole moment of the top $\hat\mu_t$, defined by the effective operator
\beq
\mathcal{L} \in - \hat\mu_t \frac{g_s}{2m_t} \bar t \sigma^{\mu\nu} T^a t G^a_{\mu\nu}~.
\eeq
The ALP contributes to $\hat\mu_t$ at one-loop order via diagrams proportional to $c^2_{tt}$ and $c_{tt}c_{GG}$, and via wavefunction renormalization of the external gluon $\propto c^2_{GG}$~\cite{Bauer:2021mvw}.
For $\Lambda_{\rm QCD} \lsim m_a < 800$~MeV we have limits from the visible mode $a \to \gamma \gamma$ at the CHARM beam dump~\cite{CHARM:1985anb}, recast by F.~Kling in Ref.~\cite{AgrawalFIPs:2021dbo}. 
Near $m_a = m_{\eta^0} \simeq 500$~MeV, due to strong ALP-$\eta^0$ mixing the ALP decay length falls short of the detector distance, resulting in a gap in the limits.

{\noindent \bf \em Supernova Limits} 

For small values of $c_{GG}\,m_n/f$ limits from data on supernovae come into play.
For $m_a$ up to around 200 MeV, the classic limit from the over-cooling of SN 1897A by the application of the so-called Raffelt criterion is shown in Fig.~\ref{fig:lims-med-ALP}.
Here we have used the limit derived in Ref.~\cite{ErtasKahlhoefer:2020xcc} assuming ALPs are produced in the supernova via bremsstrahlung in nucleon scattering due to non-zero $c_{GG}$.
In principle, ALPs are also produced by Primakoff conversion, $\gamma p \to a p$, however the limits from this process assuming non-zero $c_{\gamma \gamma}$ (trivializing all other couplings) turn out to be much weaker than those from nucleon bremsstrahlung~\cite{ErtasKahlhoefer:2020xcc}, therefore using just the latter is a good approximation here.
We also show limits from the non-observation at the SMM satellite of gamma rays produced by long-lived ALPs decaying to photons outside the proto-neutron star of SN 1987A, the so-called ``ALP burst"~\cite{Caputo:2021rux}.
Both the cooling and ALP burst limits have a ceiling from the fact that for sufficiently large couplings the ALP is trapped within the proto-neutron star. 
Finally, we also display limits recently placed in Ref.~\cite{Caputo:2022mah} by considering the boost in explosion energy from $a \to \gamma \gamma$ in several low-energy supernovae observed. 
We note that other astrophysical limits apply for coupling ranges below our description above in the $m_a \lsim 100$~MeV region~\cite{Caputo:2021rux}, but we do not display them here as the limits we show are already stronger than the NS heating sensitivities in this mass range (see Fig.~\ref{fig:lims-bauerBMcGG-masratios}).
\bigskip

{\noindent \bf \em Bounds on the UV Model} 

Our ALP model also contains heavy vector-like mediator quarks $Q$ and the radial mode $r(x)$. Both imply additional but model-dependent bounds on the theory.

With the minimal quantum numbers $\mathbf{(3,1,0)}$ we have considered so far, the $N_Q$ mediator quarks $Q$ are stable.
This makes them dangerous if they were fully thermalized in the early Universe.
Such states can annihilate very efficiently to produce small relic densities relative to dark matter for $m_Q\lesssim 10\,\tev$~\cite{Kang:2006yd,Jacoby:2007nw,DeLuca:2018mzn,Gross:2018zha}.
After QCD confinement, the heavy quarks form bound color-neutral states with themselves as well as with ordinary quarks. 
The analysis of Ref.~\cite{DeLuca:2018mzn} argues that rearrangement reactions convert the vast majority of $Q$-SM mixed states to more deeply bound, neutral $QQQ$ and $Q\bar{Q}$ modes. Despite these multiple depletion mechanisms, a small relic population of mixed $Q$-SM bound states is left over.
These carry non-integer electric charges and face extremely strong constraints of $n_{Q\text{-SM}}/n_B \lesssim 10^{-21}$ from searches for fractional charges in oil drop experiments using standard and meteoritic materials~\cite{Kim:2007zzs,Perl:2009zz}. Since we do not address the cosmological history of dark matter or the mediators in this work, we do not pursue these limits further and assume implicitly that no significant relic density of $Q$s was created.

Direct collider searches for stable $Q$ quarks place bounds that are independent of cosmology. Following $Q\bar{Q}$ pair production, the heavy quarks would be expected to hadronize and lose energy as they pass through the detector~\cite{Fairbairn:2006gg}, in analogy to long-lived $R$-hadrons considered in the context of (very) split supersymmetry~\cite{Wells:2003tf,Arkani-Hamed:2004ymt,Giudice:2004tc}. Generalizing the calculations of Ref.~\cite{Preuss:2021ebd} to color-triplet fermions, we estimate that the LHC searches of Refs.~\cite{CMS:2016kce,ATLAS:2019gqq} translate into $m_Q \gtrsim 1500\,\gev$. Note that since $m_Q/(f/c_{GG}) = y_Q\,N_Q/2\sqrt{2}$, this can be achieved for $c_{GG}/f\lesssim 10^{-2}\,\gev^{-1}$.

Collider searches also test the ALP and its related radial mode. Both produce jet final states but in different ways. For the ALP masses $m_a \leq 10\,\gev$ considered the dominant ALP decay mode is to dijets or hadrons, and the signal becomes very challenging at hadron colliders due to backgrounds. At $m_a = 10\,\gev$, the leptophobic $Z'$ search of Ref.~\cite{CMS:2019xai} translates into the bound $c_{GG}/f \lesssim 3\times 10^{-2}\,\gev^{-1}$, but we do not know of any useful bounds at lower masses.
The heavier radial mode with $m_r \sim f$ decays primarily via $r\to aa$, leading to pairs of highly boosted dijets. Assuming that each of these pairs is reconstructed as a single jet, the dijet search of Ref.~\cite{CMS:2017dcz} gives no bound for $N_Q = 1$ and $c_{GG}/f \lesssim 2\times 10^{-2}\,\gev^{-1}$ for $N_Q=10$ and $m_r=f$. We estimate that monojet-like searches for $r\to \chi\bar{\chi}$ decays such as Ref.~\cite{ATLAS:2021kxv} would give a similar sensitivity.

Motivated by the very strong bounds on stable mediator quarks, we also consider a second scenario in which they also carry hypercharge $Y=1/3$. This permits a small mass mixing with $d_R$-type quarks that allows their decays. Hypercharging the mediators also leads to couplings of the ALP and radial scalars to photons of the form
\beq
\lag_{eff}  &\supset&  \frac{\alpha}{4\pi}\,\bigg(2N_cY^2\bigg)\,c_{GG}\frac{a}{f}\,F_{\mu\nu}\widetilde{F}^{\mu\nu}
\label{eq:alpphot}
\\
&&+ \frac{\alpha}{2\pi}\,\bigg(\frac{4 N_cY^2}{3}\bigg)\,c_{GG}\,\frac{r}{f}\,F_{\mu\nu}{F}^{\mu\nu} 
\ . 
\nnmb
\eeq
For the ALP, this corresponds to $g_{a\gamma\gamma} = 2\alpha\,c_{GG}/(3\pi f) \simeq 1.7\times 10^{-3}\,c_{GG}/f$ in the standard notation~\cite{DiLuzio:2020wdo}. 

On their own, these couplings to photons (or hypercharge vector bosons more generally) do not give stronger constraints on the ALP than the gluon couplings we have considered~\cite{Bauer:2021wjo}. When both photon and gluon interactions are present, as we have here, bounds that rely on on-shell $a\to \gamma\gamma$ are weakened further by competing decays to hadrons while collider limits can become much stronger~\cite{Mariotti:2017vtv,CidVidal:2018blh,Knapen:2021elo}. Extrapolating the recent LHC boosted diphoton search of Ref.~\cite{ATLAS:2022nkn} to this scenario, we find $c_{GG}/f\lesssim 5\times 10^{-4}\,\gev^{-1}$ for $m_a = 10\,\gev$. Unfortunately, this study does not extend to lower ALP masses and therefore we do not obtain a related limit in the $m_a\in [2,10)\,\gev$ mass range of primary interest. The data scouting method proposed in Ref.~\cite{Knapen:2021elo} is projected to have sensitivity to $c_{GG}/f \sim 2\times 10^{-3}\,\gev^{-1}$. We have also studied the impact of the induced couplings of the radial mode to photons of Eq.~\eqref{eq:alpphot}, but we do not find any limits that are stronger than the dijet bounds discussed already because of the dominance of $r\to aa$ decay mode together with the very small branching ratio for $a\to \gamma\gamma$. Finally, we note that direct searches for pair-produced heavy quarks give bounds that are similar to or slightly weaker than $m_Q \gtrsim 1500\,\gev$~\cite{ATLAS:2018ziw,ATLAS:2021ibc,CMS:2022cik}.

\section{Limits on the \thdma Model\label{sec:appb}}

The \thdma model consists of a 2HDM sector together with an additional singlet pseudoscalar $a_0$. 
This latter state mixes with the $A_0$ pseudoscalar from the 2HDM to produce two physical CP-odd mass eigenstates $a$ and $A$ with mixing angle $\theta$.
In our dark matter analysis we concentrate primarily on the limit of $m_a \ll m_A$ with $|\theta| \ll 1$ such that the lighter state is mostly singlet. As a benchmark for our study, we fix $m_A = m_H = m_{H^+} = 600\,\gev$, $\tan\beta = 3$, $\lambda_3 \simeq m_h^2/v$, and $\cos(\beta-\alpha)=0$ corresponding to the so-called alignment/decoupling limit~\cite{Gunion:1989we,Djouadi:2005gj}. The dark matter dynamics then depend on $m_a$ and $\theta$, which we scan over.

These benchmark parameters fix the properties of most of the 2HDM sector, and they allow this sector to be consistent with direct and indirect bounds on it. For $\tan\beta = 3$, the heavy Higgs masses are large enough to be consistent with flavour bounds from $B\to X_s\gamma$~\cite{Misiak:2017bgg} and $B_s \to \mu\mu$~\cite{Abe:2018emu,LHCDarkMatterWorkingGroup:2018ufk}, as well as collider searches such as $H^+\to t\bar{b}$~\cite{ATLAS:2021upq} and $H/A \to \tau\bar{\tau}$~\cite{CMS:2018rmh,ATLAS:2020zms}. Furthermore, taking $m_H = m_{H^+}$ implies a residual custodial symmetry that allows the model to reproduce the precision electroweak predictions of the SM to a very good approximation~\cite{Bauer:2017ota}. Similarly, the alignment limit yields an SM-like Higgs boson with production and decay rates effectively identical to the SM~\cite{Gunion:1989we,Djouadi:2005gj}. 

While the 2HDM sector is largely safe in this benchmark, the presence of a lighter pseudoscalar implies bounds on the mixing angle $|\theta|$ for a given mass $m_a$. As mentioned in Sec.~\ref{sec:2HDMa_limits}, flavor bounds can be important in this benchmark as well for $m_a\lesssim 10\,\gev$. In our region of parameter space, these are dominated by the flavor-changing coupling of $a$ to a bottom and strange quark induced at one-loop~\cite{Arcadi:2017wqi}. Using the results of~\cite{Arnan:2017lxi}, we compute the rate for $B^\pm\to{K^\pm}^{(\ast)}(a\to\mu^+\mu^-)$ and compare to the LHCb search for promptly-decaying bosons in this channel~\cite{LHCb:2015nkv}. In addition, there is a subdominant constraint that comes from requiring that the branching fraction for $B_s\to\mu^+\mu^-$ agree with the world average of $(3.01\pm0.35)\times10^{-9}$~\cite{ParticleDataGroup:2020ssz}. The resulting limits on $\theta$ from these processes are shown in Fig.~\ref{fig:limsmed-2hdma}.

Pseudoscalar masses $m_a \lesssim m_h/2$ are also strongly constrained by searches for rare SM Higgs decays. The decay width for $h\to aa$ in our benchmark is~\cite{Ipek:2014gua,Bauer:2017ota,Abe:2018emu}
\beq
\Gamma(h\to aa) = \frac{\lambda_{haa}^2}{32\pi}\,m_h\,\sqrt{1-(2m_a/m_h)^2} \ ,
\eeq
where
\beq
\lambda_{haa} = \frac{(m_A^2-m_a^2)}{2m_h\,v}\,\sin^22\theta \ .
\eeq
With $m_A = 600\,\gev$ this width exceeds the total SM Higgs width for $|\theta| \gtrsim 0.05$ and thus Higgs measurements imply bounds on the mixing angle near this level. Since we focus on $m_a \leq m_{\chi}$, the $a$ pseudoscalar decays visibly through its mixing with the 2HDM pseudoscalar $A$. Expressions for the widths of the $a$ state are collected in Refs.~\cite{Bauer:2017ota,Haisch:2018kqx}. Exotic decays of the SM Higgs have been searched for the by LHC collaborations~\cite{Cepeda:2021rql}, and in Fig.~\ref{fig:limsmed-2hdma} we show estimated limits from ATLAS searches for $h\to aa \to bb\mu\mu$~\cite{ATLAS:2021hbr} and $h\to aa\to \tau\tau\mu\mu$~\cite{ATLAS:2015unc} as well as CMS searches for $h\to aa\to bb\tau\tau$~\cite{CMS:2018zvv} and $h\to aa\to 4\mu$~\cite{CMS:2018jid}. Non-standard Higgs decays are also constrained by Higgs rate measurements if they do not contribute to the standard SM Higgs search channels. Global fits to ATLAS Higgs measurements put a limit on \emph{undetected} Higgs decays of BR$_\text{und} < 0.15$~\cite{ATLAS:2019nkf,ATLAS:2021vrm}. We also show the corresponding limit on $|\theta|\lesssim 0.04$ in Fig.~\ref{fig:limsmed-2hdma} under the assumptions of unmodified SM Higgs production and that $h\to aa$ decays do not significantly pollute SM Higgs rates.

Direct searches at the LHC become more important for pseudoscalar masses above the Higgs decay threshold. Collider bounds on the \thdma model have been studied extensively for $m_a > 2\,m_{\chi}$ with dominant $a\to \chi\bar{\chi}$ decays~\cite{No:2015xqa,Arcadi:2017wqi,Haisch:2018kqx,LHCDarkMatterWorkingGroup:2018ufk}, but they have not been investigated in as much detail when this channel is closed and the pseudoscalar decays visibly. We find that LHC searches for $H/A\to \tau\tau$ can be reinterpreted to give the strongest limits on this scenario. For our benchmark of $\tan\beta = 3$, the most important production channel is in association with bottom quarks. We estimate the production cross section by rescaling the $bb\phi$ cross sections tabulated in Ref.~\cite{LHCHiggsCrossSectionWorkingGroup:2016ypw} by the appropriate factor of $(\tan\beta\,\sin\theta)^2$, and derive bounds on $|\theta|$ by comparing to the limits obtained from $\phi\to \tau\tau$ searches at ATLAS~\cite{ATLAS:2020zms} and CMS~\cite{CMS:2018rmh}. The exclusions we find based on these studies are shown in Fig.~\ref{fig:limsmed-2hdma}.

\begin{figure*}[t]
    \centering
    \includegraphics[width=.49\textwidth]{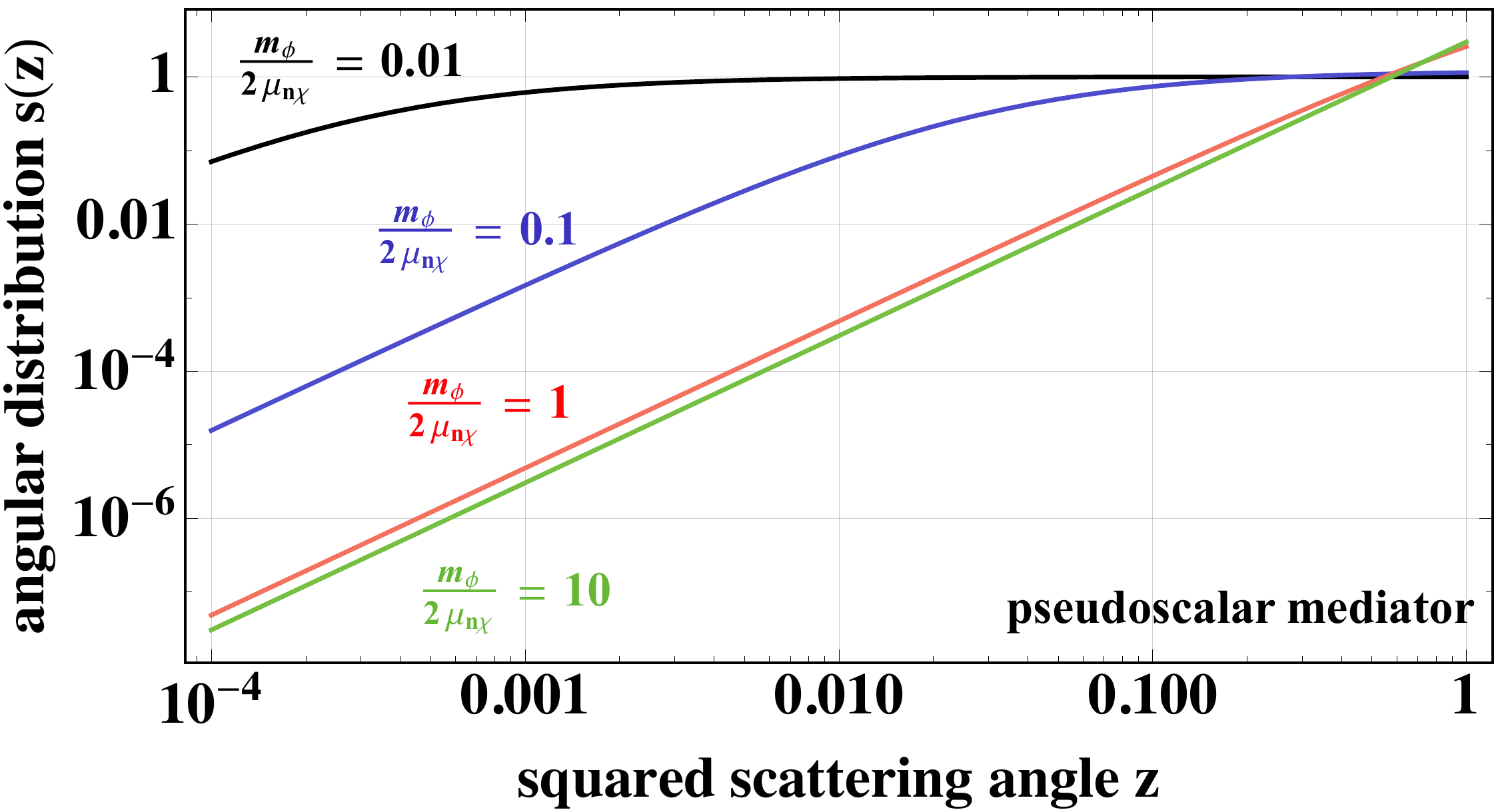} \
    \includegraphics[width=.49\textwidth]{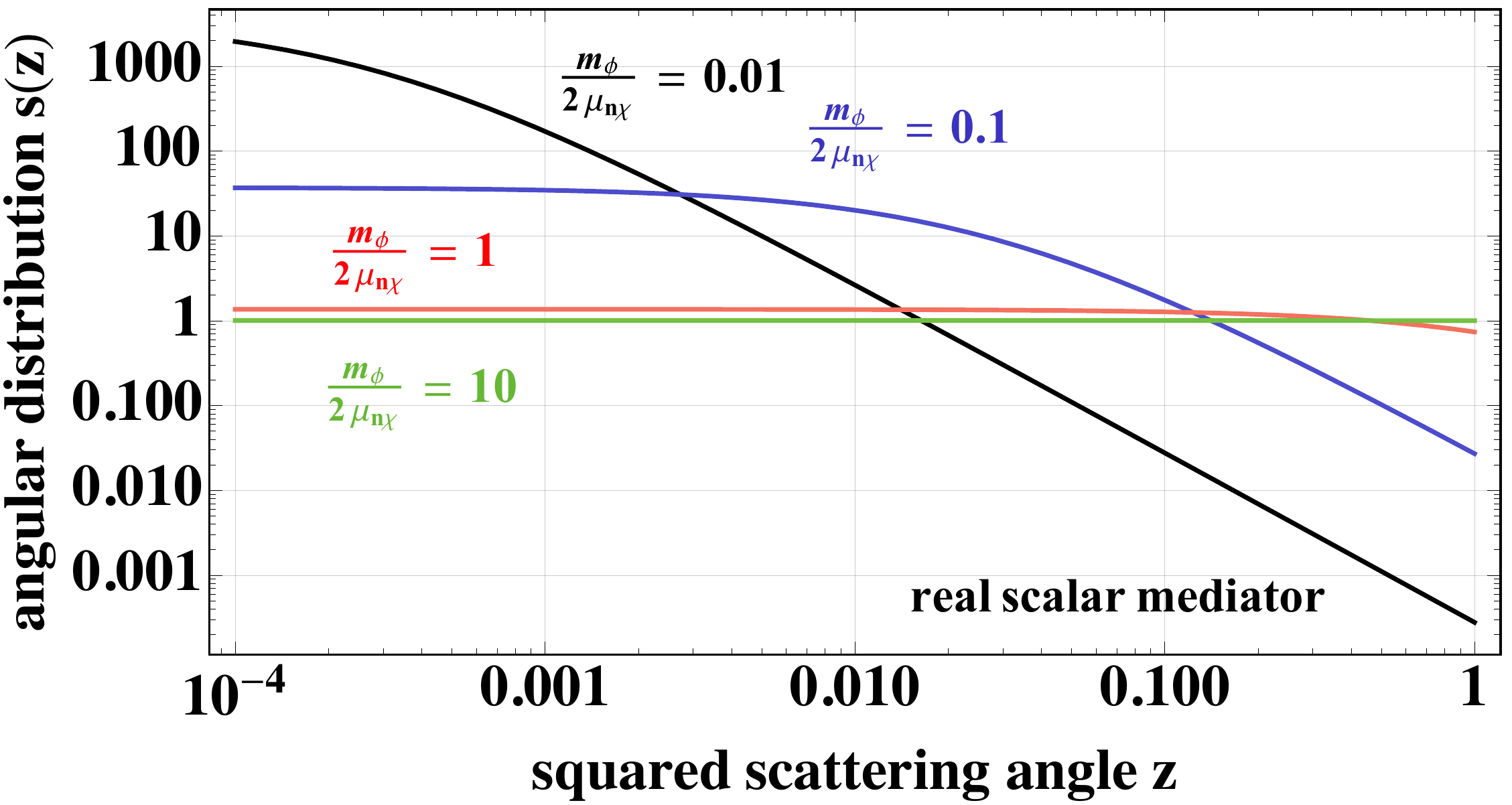} \\ 
    \includegraphics[width=.49\textwidth]{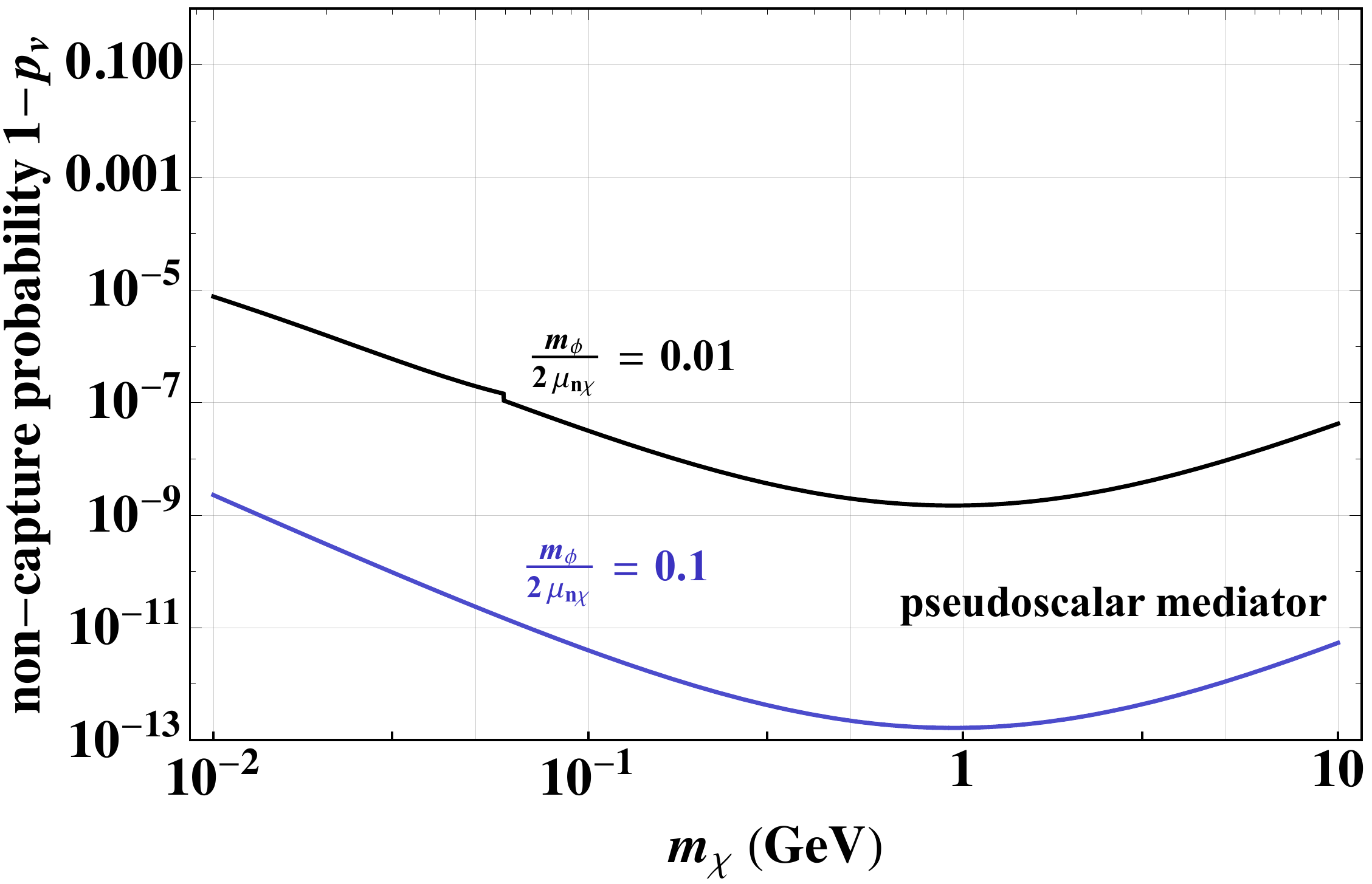} \
     \includegraphics[width=.49\textwidth]{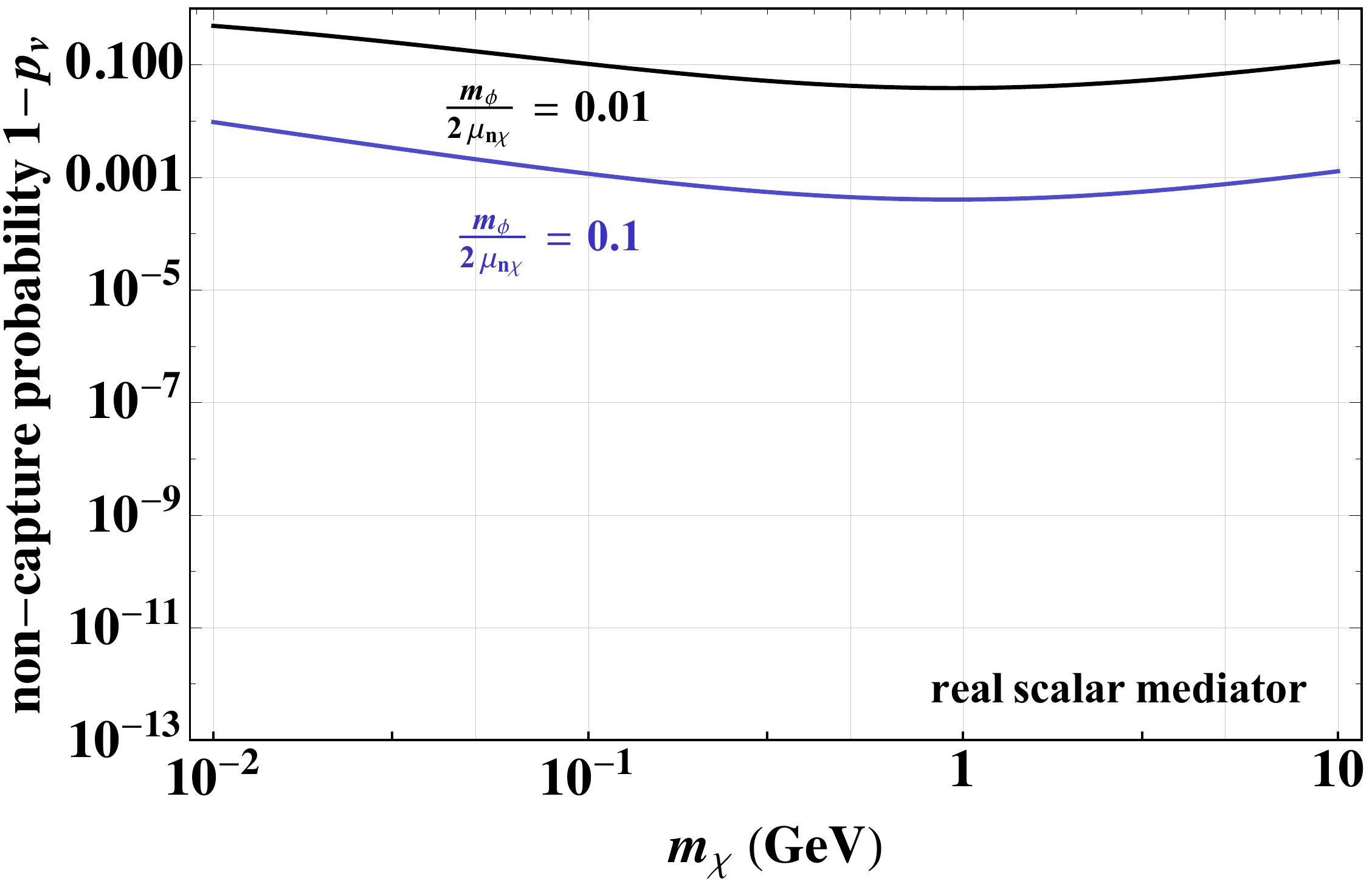} 
    \caption{\raggedright{\small {\bf \em Top}: Event distributions with respect to DM-nucleon scattering angle variable $z \equiv \cos^2 \theta_{\rm lab}$, {\bf \em bottom}: probability of not capturing in the NS following DM-nucleon scattering, {\bf \em left}: pseudoscalar mediation, {\bf \em right}: real scalar mediation.
    Due to the $q^4$ dependence in the cross section, the case of pseudoscalar mediators does not prefer forward scatters with low energy transfers, unlike light scalar mediators.
    This results in almost all scatters resulting in capture.
    See Appendix~\ref{app:lightmedNS} for further details.
    }
     }
    \label{fig:Elossdis}
\end{figure*}

\section{Light Mediator Kinematics\\ and NS Capture\label{sec:appc}}
\label{app:lightmedNS}

In this appendix we show that the fraction of DM scattering events in the NS that do not result in capture is highly negligible in our scenario, i.e. $p_v \to 1$ in Eq.~\eqref{eq:masscaprate}.
The case of $p_v \ll 1$ could potentially occur if the mediator is much lighter than the momentum transfer, favoring forward scatters with small energy losses insufficient for capture.
This can be seen in the normalized event distribution with respect to the scattering angle variable $z \equiv \cos^2\theta_{\rm lab} = (1-\cos \theta_{\rm CM})/2$,
\beq
s(z) \equiv \frac{1}{\sigma}\frac{d\sigma}{dz}~,
\eeq
plotted in the top panels of Fig.~\ref{fig:Elossdis} with the left (right) panel corresponding to mediation by a CP-odd (CP-even) scalar of mass $m_\phi$.
Here we show distributions for various values of $m_\phi/(2\mu_{n\chi})$ scanning between the ``light" and ``heavy" mediator regimes, since the momentum transfer $Q \simeq 2 \mu_{n\chi} \cos\theta_{\rm lab}$.
We find agreement for the CP-even scalar case with Ref.~\cite{NSvIR:DasguptaGuptaRay:LightMed}.

We see that real scalar mediation is biased toward forward events with small $z$ as the mediator gets lighter,
as expected for $d\sigma/dz \propto (Q^2+m_\phi^2)^{-2}$.
However, for pseudoscalar mediation we see that events are biased toward {\em larger} $z$ as the mediator gets lighter, with $s(z)$ getting flatter.
This is because here $d\sigma/dz \propto Q^4/(Q^2+m_\phi^2)^{2}$; as $m_\phi$ is decreased, backward events are more favored, and the sensitivity to $z$ is lost for $m_\phi \ll q$.

Capture in the NS is guaranteed if scattering on its constituents depletes DM of its halo kinetic energy.
Thus the probability of post-scattering capture is defined by~\cite{NSvIR:DasguptaGuptaRay:LightMed}
\bea
\nn p_v &=& \int_{z_{\rm min}}^1 dz \ s(z)~,\\
z_{\rm min} &=& \frac{(\mdm + m_n)^2}{4\mdm m_n} \frac{v_{\rm halo}^2}{v_{\rm halo}^2 + v_{\rm esc}^2}~,
\eea
where we set $v_{\rm halo} = 10^{-3} c$ and $v_{\rm esc} = 0.59 c$ in practice.
In the bottom panels of Fig.~\ref{fig:Elossdis} we plot the probability of not capturing, $1-p_v$, for both the CP-odd and CP-even mediators in the light regime.
As a result of events weighted toward $z = 1$, we find $1 - p_v$ negligible, i.e. $p_v = 1$ is an excellent approximation in our treatment.
This is not the case for CP-even scalar mediators, where we see that it is possible for an $\Oc(1)$ fraction of scattering events to evade capture for $m_\phi/(2\mu_{\rm n\chi}) \leq 0.01$, as also shown in Ref.~\cite{NSvIR:DasguptaGuptaRay:LightMed}.

\bibliography{refs}

\end{document}